\documentclass[a4paper,11pt]{article}
\usepackage{jcappub}
\usepackage[T1]{fontenc}
\usepackage[utf8]{inputenc}
\usepackage{units}
\usepackage{url}

\title{Indirect detection constraints on the scotogenic dark matter model}

\author[a]{T. de Boer,}
\author[b]{R. Busse,}
\author[b]{A. Kappes,}
\author[a,1]{M. Klasen,\note{michael.klasen@uni-muenster.de}}
\author[a]{and S. Zeinstra}

\affiliation[a]{Institut für Theoretische Physik, Westfälische Wilhelms-Universität Münster,\\Wilhelm-Klemm-Str. 9, 48149 Münster, Germany}
\affiliation[b]{Institut für Kernphysik, Westfälische Wilhelms-Universität Münster,\\Wilhelm-Klemm-Str. 9, 48149 Münster, Germany}

\abstract{
  Radiative seesaw models have the attractive property of providing dark matter candidates in addition to the generation of neutrino masses. Here we present a study of neutrino signals from the annihilation of dark matter particles that have been gravitationally captured in the Sun in the framework of the scotogenic model. We compute expected event rates in the \textsc{IceCube} detector in its 86-string configuration. As fermionic dark matter does not scatter off nucleons due to its singlet nature and therefore does not accumulate in the Sun, we study the case of scalar dark matter with a scan over the parameter space. Due to a naturally small mass splitting between the two neutral scalar components, inelastic scattering processes with nucleons can occur. We find that for most of the parameter space, i.e.\ for mass splittings below 500 keV, inelastic scattering in the Sun yields \textsc{IceCube} event rates above $10$ events per year, whereas direct detection on Earth is sensitive only to 250 keV. Consequently, a detailed analysis with \textsc{IceCube} could lead to a lower limit on the scalar coupling $\lambda_5\gtrsim1.6\cdot10^{-5}\cdot m_{\rm DM}/$TeV. For larger mass splittings, only elastic scattering occurs in the Sun. In this case, \textsc{XENON1T} limits only allow for models with expected event rates of up to $\mathcal{O}(0.1)$ per year. Some of these models, in particular those with large DM mass and fermion coannihilation, could also be tested with a dedicated \textsc{IceCube} analysis of DM annihilation in the Galactic Center.
}

\begin{document}
\maketitle
\flushbottom

\section{Introduction}

While there is today overwhelming evidence for dark matter (DM) in the Universe from observations at many different length scales \cite{Zyla:2020zbs,Klasen:2015uma}, and its relic density can be precisely determined \cite{Agha18}, the nature of DM remains still unknown. Among the many theoretical ideas that have been put forward to explain it, radiative seesaw models \cite{Restrepo:2013aga,Klasen:2016vgl,Esch:2018ccs,Fiaschi:2018rky,deBoer:2021xjs} such as the famous scotogenic model \cite{Ma06,KRRYZ13,Toma:2013zsa,Ibarra:2016dlb,BKRZ20} have the advantage of extending the Standard Model (SM) with only very few fields. They include not only a natural DM candidate, but also explain the smallness of neutrino masses and thus address simultaneously two of the largest current mysteries of theoretical particle physics.

In radiative seesaw models, DM in the form of Weakly Interacting Massive Particles (WIMPs) can annihilate either directly into neutrinos or via the decays of other intermediately produced SM particles. Previous work focused on monochromatic neutrinos from direct decays, as they are easier to distinguish from the background \cite{Lindner:2010rr,Farzan:2011ck,Arina:2015zoa,EGHV17}. Here, we consider both direct as well as secondary neutrinos from the decays of intermediate other SM particles.

In order to boost the amount of WIMP annihilations, one can consider regions with a local overdensity \cite{JUNGMAN1996195}. Since our solar system is embedded in the galactic DM halo, WIMPs can accumulate in large celestial bodies like the Sun, which we focus on in this work. Upon scattering with a nucleus inside the Sun, a WIMP can lose enough kinetic energy to be captured by the Sun's gravitational potential. Thus the WIMP-nucleon scattering cross section plays an important role in this capture process. We go beyond the standard scenario of elastic DM-nucleon scattering by considering also inelastic scattering processes, in which a WIMP up-scatters to a slightly heavier state. This so-called inelastic DM had originally been proposed by Smith and Weiner \cite{TuckerSmith:2001hy} to explain the annual modulation signal at \textsc{DAMA/LIBRA} \cite{Bernabei:2018jrt,Bernabei:2019ajy}. While these authors considered the sneutrino as a specific DM candidate, they concluded generally that due to the larger DM velocity, the inelasticity is in fact less relevant in the Sun than at direct detection experiments, leaving ample room for indirect detection experiments.

Other previous work that considered the prospect of detecting inelastic DM indirectly includes Refs.\ \cite{Nussinov:2009ft,Menon:2009qj,Shu:2010ta}. This work was motivated by the \textsc{DAMA/LIBRA} signal and principally considered the parameter space that fitted this signal. A comparison between \textsc{IceCube} and direct detection experiments in a more general inelastic scenario has been carried out within the context of effective field theory in Ref.\ \cite{Catena:2018vzc}. There it was found that neutrino telescopes should place stronger limits than direct detection experiments for mass splittings larger than about 200 keV for DM particles of mass 1 TeV.

Of the many models that connect neutrinos and DM, the scotogenic model is the best-known example \cite{Ma06}. Its main strength is the relatively simple extension of the SM with only two new fields, whilst still providing enough interesting phenomenology, in particular naturally occurring inelastic DM. The neutrino masses are generated at loop level through the radiative seesaw mechanism, whereas the tree level seesaw is forbidden by a $\mathbb{Z}_2$ symmetry under which all new particles have an odd charge. In this work we focus on the scotogenic model with scalar DM. For a similar model, extended by a real scalar singlet to account for inflation, elastic scattering in the Sun was found to produce neutrino signals at least two orders of magnitude below the current sensitivity of neutrino telescopes \cite{Hashimoto:2020xoz}.

As the largest neutrino telescope worldwide, the \textsc{IceCube} Observatory \cite{Aart16} is predestined to search for neutrino signals from annihilating WIMPs and thereby contribute to the search for physics beyond the SM. We investigate the parameter space of the scotogenic model for a detectable neutrino flux in \textsc{IceCube} from WIMP annihilations in the Sun and Galactic Center, which could therefore be used to constrain the scotogenic parameter space with dedicated \textsc{IceCube} data analyses. 

This work is organised as follows: We introduce the scotogenic model in chapter \ref{sec: The model}, and discuss scattering processes and DM capture in chapter \ref{sec: Capture}. In chapter \ref{sec: Detection IceCube}, we explain the detection of neutrinos from WIMP annihilations in the Sun with \textsc{IceCube}, where we show observables of a benchmark point. The results of the scans over the parameter space are discussed and compared to experimental limits in chapter \ref{sec: Parameter scans}. We summarize our findings and present an outlook to future studies in chapter \ref{sec: Summary}.

\section{The scotogenic model\label{sec: The model}}

The scotogenic model considered in this work extends the SM by two new fields, a scalar doublet with the components $\left(\eta^{+},\eta^{0}\right)$ and three generations of a fermion singlet $N_{i}$, where $i=1,2,3$. In addition to the SM gauge group, the model assumes a discrete $\mathbb{Z}_{2}$ symmetry, under which the new fields are all odd and the SM fields are even, which prevents further decay of the lightest new mass eigenstate into SM particles and thus guarantees, if this eigenstate is neutral, the stability of the DM candidate \cite{Ma06}.

Apart from the usual kinetic terms, the new terms in the Lagrangian are
\begin{eqnarray}
\mathcal{L}_{N} & = & -\frac{m_{N_{i}}}{2}N_{i}N_{i}+y_{i\alpha}\left(\eta^{\dagger}L_{\alpha}\right)N_{i}+\text{h.c.}-V,\label{eq: lagrangian}
\end{eqnarray}
where $m_{N_{i}}$ is the (diagonal) mass matrix of the fermion singlets. The $L_{\alpha}$ denote the three generations of the left-handed SM lepton doublets ($\alpha=1,2,3$), and $y_{i\alpha}$ is a $3\times3$ Yukawa coupling matrix.

The scalar potential is given by
\begin{eqnarray}
V & = & m_{\phi}^{2}\phi^{\dagger}\phi+m_{\eta}^{2}\eta^{\dagger}\eta+\frac{\lambda_{1}}{2}\left(\phi^{\dagger}\phi\right)^{2}+\frac{\lambda_{2}}{2}\left(\eta^{\dagger}\eta\right)^{2}+\lambda_{3}\left(\phi^{\dagger}\phi\right)\left(\eta^{\dagger}\eta\right)\nonumber \\
 &  & +\lambda_{4}\left(\phi^{\dagger}\eta\right)\left(\eta^{\dagger}\phi\right)+\frac{\lambda_{5}}{2}\left[\left(\phi^{\dagger}\eta\right)^{2}+\left(\eta^{\dagger}\phi\right)^{2}\right],\label{eq: lagrangian potential}
\end{eqnarray}
where $\phi$ is the SM Higgs field with vacuum expectation value $\langle\phi^{0}\rangle=246.22$ GeV$/\sqrt{2}$ \cite{Zyla:2020zbs}. Vacuum stability requires the scalar couplings to obey the relations \cite{LPYM16}
\begin{align}
\lambda_{1}>0,&\quad \lambda_{2}>0,\nonumber\\
\lambda_{3}&>-\sqrt{\lambda_{1}\lambda_{2}}, \\
\lambda_{3}+\lambda_{4}-\left|\lambda_{5}\right|&>-\sqrt{\lambda_{1}\lambda_{2}} \nonumber,\label{eq: vacuum stability}
\end{align}
while perturbativity imposes $|\lambda_i|<4\pi$. After electroweak symmetry breaking, the new Lagrangian gives rise to three physical scalar bosons with squared masses
\begin{eqnarray}
m_{\eta^{+}}^{2} & = & m_{\eta}^{2}+\lambda_{3}\langle \phi^{0}\rangle^{2},\nonumber \\
m_{\eta^{0R}}^{2} & = & m_{\eta}^{2}+\left(\lambda_{3}+\lambda_{4}+\lambda_{5}\right)\langle \phi^{0}\rangle^{2},\label{eq: scalar masses}\\
m_{\eta^{0I}}^{2} & = & m_{\eta}^{2}+\left(\lambda_{3}+\lambda_{4}-\lambda_{5}\right)\langle \phi^{0}\rangle^{2}.\nonumber 
\end{eqnarray}
The measurement of the SM Higgs boson mass $m_h$ at the LHC \cite{Aad15} and the relation $m_{h}^{2}=2\lambda_{1}\langle\phi^{0}\rangle^{2}=-2m_\phi^2=(125$\ GeV$)^2$ fix $\lambda_1$ to $0.26$. Since $\lambda_{2}$ induces only self-interactions of the new scalars, which do not affect the phenomenology, we set it to $\lambda_{2}=0.5$ without loss of generality. We assume DM to be the lightest neutral scalar, i.e.\ either the real or the imaginary component of $\eta^0=(\eta^{0R}+i\eta^{0I})/\sqrt{2}$. Their mass splitting, induced by the coupling $\lambda_5$, is naturally small, since if $\lambda_5$ is exactly zero, the neutrinos would be massless and lepton number would be conserved, leading to a larger symmetry of the Lagrangian. This and the fact that the lightest scalar must be neutral imply that $\lambda_4<0$ \cite{Hashimoto:2020xoz}. When $m_\eta^2$ dominates over $\langle\phi_0\rangle^2$, the scalar couplings $\lambda_3$ and $\lambda_4$ will play a subdominant role in the mass splitting and $\eta^+$ will be close in mass to both $\eta^{0R}$ and $\eta^{0I}$.

\begin{figure}
  \centering
  \includegraphics[width=0.4\textwidth]{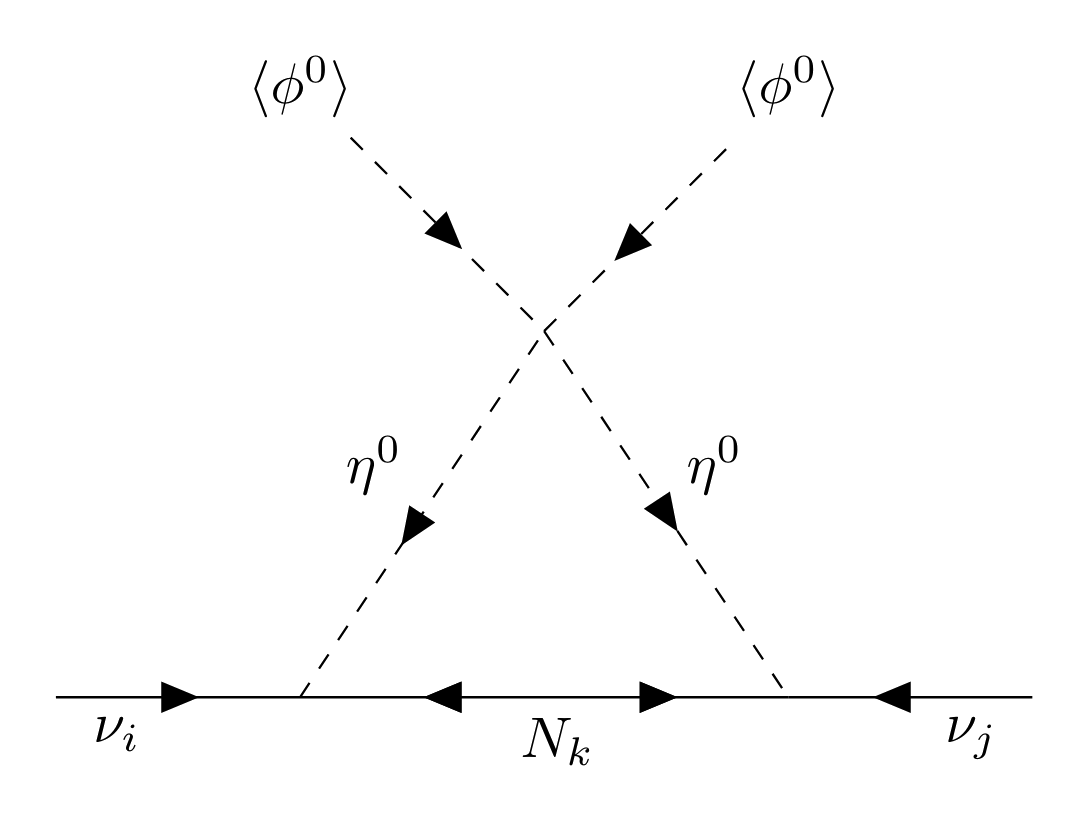}
  \caption{\label{diagram: scotogenic one-loop neutrino masses}One loop generation of neutrino masses in the scotogenic model.}
\end{figure}

In the scotogenic model, the SM neutrinos obtain their masses through the one loop diagram shown in Fig.\ \ref{diagram: scotogenic one-loop neutrino masses} \cite{Ma06}.
The SM neutrino mass matrix is connected to the masses of the dark sector fields by
\begin{equation}
\left(m_{\nu}\right)_{\alpha\beta}=\left(y^{T}\Lambda y\right)_{\alpha\beta}\label{eq: SM neutrino mass matrix}
\end{equation}
with the Yukawa matrix $y$ and the diagonal mass matrix $\Lambda$, which has the eigenvalues 
\begin{equation}
\Lambda_{i}=\frac{m_{N_{i}}}{32\pi^{2}}\left[\frac{m_{\eta^{0R}}^{2}}{m_{\eta^{0R}}^{2}-m_{N_{i}}^{2}}\log\left(\frac{m_{\eta^{0R}}^{2}}{m_{N_{i}}^{2}}\right)-\left(R\rightarrow I\right)\right]\,.\label{eq: Diagonal mass matrix eigenvalues}
\end{equation}
The two terms within the brackets are equal up to interchanged real and imaginary components. As explained above, non-zero neutrino masses therefore require a mass splitting between $m_{\eta^{0R}}$ and $m_{\eta^{0I}}$. The PMNS matrix $U_{\text{PMNS}}$ diagonalizes the neutrino mass matrix with
\begin{equation}
U_{\text{PMNS}}^{T}m_{\nu}U_{\text{PMNS}}=\hat{m}\equiv\text{diag}\left(m_{1},m_{2},m_{3}\right)\,.
\end{equation}
For given masses and scalar couplings of the model, the Yukawa matrix
\begin{equation}
y=\sqrt{\Lambda}^{-1}R\sqrt{\hat{m}_{\nu}}U_{\text{PMNS}}^{\dagger}\,,\label{eq: Casas-Ibarra}
\end{equation}
is then fixed by the known SM neutrino mass differences and mixing angles through the Casas-Ibarra parametrization up to an orthogonal matrix $R$, that depends on three arbitrary rotation angles $\theta_{i}\in\left[0;2\pi\right]$ \cite{CI01}.

To obtain the correct DM relic density measured by \textsc{Planck} \cite{Agha18}, not only annihilation processes of the lightest scalar, but also coannihilation processes with one or several dark fermions can be very important \cite{KRRYZ13}. Since the SM neutrinos are part of a lepton doublet, charged lepton flavor violation (LFV) naturally occurs in radiative seesaw models. In the scotogenic model, the corresponding constraints have a larger impact on fermion than on scalar DM, which we consider here \cite{Toma:2013zsa}. Nevertheless, we take the LFV limits into account as discussed below. Conversely, direct detection has a large impact on scalar DM, as we will also see below, whereas fermion DM us only impacted at one loop \cite{Ibarra:2016dlb}. The LEP2 limit on new charged scalar masses ($m>98.5$ GeV) \cite{Abbi03} is usually avoided due to either the dominance of $m_\eta^2$ over $\langle\phi_0\rangle^2$ or the large splitting with the neutral scalar masses induced by $\lambda_4$. Note that in addition the parameter space of fermion DM is further constrained by the \textsc{KATRIN} upper limit on the lightest neutrino mass \cite{BKRZ20}.

\section{WIMP capture in the Sun}\label{sec: Capture}

When WIMPs scatter off nuclei in the Sun, they can lose energy up to a point where their remaining kinetic energy is insufficient to escape the gravitational potential of the Sun. This leads to an accumulation of DM in the Sun's core, boosting the annihilation rate. The neutrino flux from these annihilations is thus characterized by the WIMP-nucleon scattering cross section. As described above, the scotogenic model can provide either a fermionic ($N_{i}$) or scalar ($\eta^{0R}$, $\eta^{0I}$) DM candidate, whichever is the lightest particle of the $\mathbb{Z}_2$ odd sector.

Singlet fermion DM does not scatter off nucleons at tree level. It therefore does not accumulate in the Sun (or the Earth) and produce a significant neutrino flux, rendering it undetectable for neutrino telescopes. In addition, only direct annihilations into SM neutrinos and charged leptons are possible, which are both suppressed by small Yukawa couplings and heavy scalar propagators. This also holds for gravitationally accumulated fermion singlet DM in the Galactic Center. Scalar doublet DM, on the other hand, has electroweak size cross sections, which enable accumulation in celestial bodies and annihilation into SM particles, that subsequently decay into neutrinos. In this work we therefore only consider scalar DM, as it can produce sizable detection rates at neutrino telescopes.

\begin{figure}
  \centering
  \includegraphics[width=0.5\textwidth]{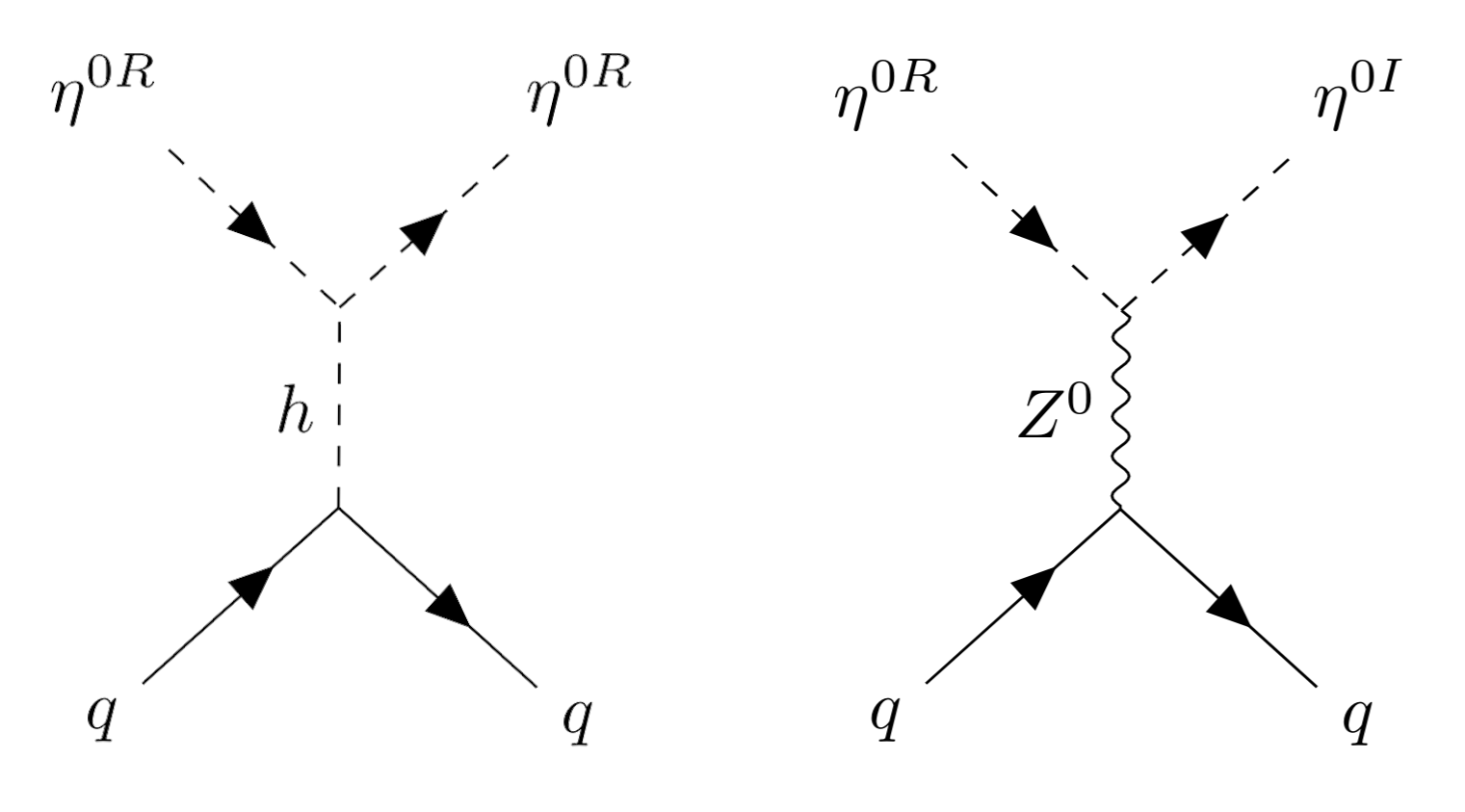}
  \caption{\label{diagram: scotogenic SISD scattering}Feynman diagrams of the elastic (left) and inelastic (right) scalar DM-nucleon scattering processes in the scotogenic model. If $\eta^{0I}$ is the DM candidate, $\eta^{0R}$ and $\eta^{0I}$ change their roles. For mass splittings larger than a few hundred keV, the right diagram is kinematically forbidden.}
\end{figure}

The diagrams for scalar DM scattering off nucleons are shown in Fig. \ref{diagram: scotogenic SISD scattering}.
Usually these scattering processes are described by diagrams with the same in- and outgoing DM particle. This is the elastic case. However, as originally pointed out in Ref.\ \cite{TuckerSmith:2001hy}, the existence of a slightly heavier state allows also for inelastic upscattering of the DM particle, provided that the mass splitting between the two states $\delta=|m_{\eta^{0R}}-m_{\eta^{0I}}|$ fulfills
\begin{equation}
    \delta < \frac{\mu v^2}{2},
\end{equation}
where $\mu$ is the WIMP nucleus reduced mass and $v$ is the relative velocity. In the scotogenic model, the mass splitting between the neutral scalar components is governed by $\lambda_5$. Since $\lambda_5$ is naturally small as explained in Sec.\ \ref{sec: The model}, the mass splitting is small as well and can be approximated by
\begin{equation}
    \delta \approx \frac{\lambda_5 \langle\phi^{0}\rangle^2}{m_{\eta^{0R,I}}}.
    \label{eq: del lam5}
\end{equation}

\subsection{Elastic scattering}

First we turn our attention to the case of elastic scattering. As shown in Fig. \ref{diagram: scotogenic SISD scattering} (left), the elastic process is mediated by the exchange of the SM Higgs boson $h$. The Lagrangian relevant for this interaction is given by \cite{Agrawal:2010fh} 
\begin{align}
  \mathcal{L}=-\frac{1}{2}g_{\eta^{0R}\eta^{0R}h}\left(\eta^{0R}\right)^2h - g_{\bar{q}qh}\bar{q}qh,
\end{align}
assuming that $\eta^{0R}$ is the dark matter candidate. When $\eta^{0I}$ is the lightest dark particle, one simply needs to replace $\eta^{0R}$ with $\eta^{0I}$. The couplings $g_{\eta^{0R,I}\eta^{0R,I}h}=(\lambda_3+\lambda_4\pm\lambda_5)\langle\phi^{0}\rangle\sqrt{2}$ and $g_{\bar{q}qh}=y_q/\sqrt{2}$ are fixed by the high energy Lagrangian of the scotogenic model in Sec.\ \ref{sec: The model} and the SM quark Yukawa couplings $y_q=m_q/\langle\phi^{0}\rangle$ and can be obtained using \textsc{SARAH} \cite{Staub:2013tta}.

In the non-relativistic limit, an effective Lagrangian can be derived by integrating out the mediator. The result is \cite{Agrawal:2010fh,JUNGMAN1996195} 
\begin{align}
	\mathcal{L}_\text{eff}=\frac{1}{2}a_q2m_{\eta^{0R}}\left(\eta^{0R}\right)^2\bar{q}q,
\end{align}
where we have grouped a factor $2m_{\eta^{0R}}$ with the effective operator such that it has the same mass dimension as in the fermionic case \cite{Freytsis:2010ne}. The effective coupling is thus defined as $a_q:=g_{\eta^{0R,I}\eta^{0R,I}h}g_{\bar{q}qh}/(2m_{\eta^{0R,I}}m_h^2)$ and has mass dimension $(-2)$ as usual. The elastic scattering amplitude on a nucleon $N$, i.e.\ a proton $p$ or a neutron $n$,  is then given by 
\begin{align}
  \mathcal{M}=4m_{\eta^{0R}}m_N\sum_qa_q\langle N|\bar{q}q|N\rangle,
\end{align}
where $m_N$ is the nucleon mass and $|N\rangle$ is the nucleon state in the non relativistic normalization.

The nucleon matrix elements for light quarks ($q=u,d,s$) can be evaluated as \cite{Agrawal:2010fh,JUNGMAN1996195} 
\begin{align}
	m_{q}\langle N|\bar{q}q|N\rangle=m_Nf_q^{(N)},
\end{align}
where $m_q$ are the quark masses and $f_q^{(N)}$ denote the scalar quark form factors in nucleons, for which we employ the values shown in Tab. \ref{tab:nuclear_FF} \cite{BBGPZ18}.
\begin{table}
\caption{Scalar quark form factors in nucleons used in this work \cite{BBGPZ18}.}
\label{tab:nuclear_FF}
\centering
\begin{tabular}{|c|ll|}
		\hline
		&proton&neutron\\
		\hline
 		$f^{(N)}_d$&  0.0191  &  0.0273 \\
  		$f^{(N)}_u$&  0.0153  &  0.011  \\
  		$f^{(N)}_s$&  0.0447  &  0.0447 \\
		\hline
\end{tabular}
\end{table}
For heavy quarks ($Q=c,b,t$), the matrix element is given by 
\begin{align}
	m_{Q}\langle N|\bar{Q}Q|N\rangle=\frac{2}{27}m_N\left[1-\sum_{q=u,d,s}f_q^{(N)}\right].
\end{align}
The effective coupling to protons (neutrons) is then 
\begin{align}
	\frac{f_{p(n)}}{m_{p(n)}}=\sum_{q=u,d,s}f_q^{p(n)}\frac{a_q}{m_q}+\frac{2}{27}\left[1-\sum_{q=u,d,s}f_q^{p(n)}\right]\sum_{Q=c,b,t}\frac{a_{Q}}{m_{Q}}.
\end{align}

Spin independent interactions couple coherently to the $Z$ protons and $A-Z$ neutrons in a nucleus of mass number $A$ and mass $m_A$. At zero momentum transfer $|{\bf q}|$, the scattering cross section can therefore be written as \cite{Agrawal:2010fh,JUNGMAN1996195} 
\begin{align}
  \sigma_A^0=\frac{\mu^2}{\pi}\left[Zf_p+(A-Z)f_n\right]^2.
  \label{eq: sigma0}
\end{align}
Taking into account the loss of coherence at high momentum transfer, the differential cross section 
\begin{align}
  \frac{d\sigma_A}{d|{\bf q}|^2}=\frac{\sigma_A^0}{4\mu^2v^2}F^2(Q)
\end{align}
depends also on a nuclear form factor $F^2(Q)$, where the energy transfer $Q$ (or recoil energy $E_R$) is $|{\bf q}|^2/(2m_A)$ and ranges from $Q_\text{min}=0$ to $Q_\text{max}=4\mu^2v^2/(2m_A)$. In this work, we use the Gaussian form factor \cite{Gould:1987} 
\begin{align}
  F^2(Q)=\exp(-Q/Q_0)
\end{align}
with $Q_0=3/(2m_AR^2)$ and 
\begin{align}
  R=\left[0.91\left(\frac{m_A}{\text{GeV}}\right)^{\frac{1}{3}}+0.3\right]\times10^{-13}\text{ cm},
\end{align}
since it allows for analytic integration of the scattering cross section and thus for a fast subsequent numerical evaluation of the neutrino flux in large scans of the model parameter space.

\subsection{Inelastic scattering}

The inelastic scattering of scalar DM off nuclei in the scotogenic model is mediated by the exchange of a $Z^0$ boson as shown in Fig.\ \ref{diagram: scotogenic SISD scattering} (right). The vector part of the Lagrangian coupling neutral scalars and quarks to the $Z^0$ boson is given by \cite{Agrawal:2010fh} 
\begin{align}
	\mathcal{L}=g_{\eta^{0R}\eta^{0I}Z^0}\left(\eta^{0R}\partial^\mu\eta^{0I}-\partial^\mu\eta^{0R}\eta^{0I}\right)Z^0_\mu +g_{\bar{q}qZ^0}\bar{q}\gamma^\mu qZ^0_\mu.
\end{align}
Note that elastic scattering is absent for the $Z^0$ boson and that the axial vector part of the $\bar{q}qZ^0$ interaction vanishes in the non relativistic limit. The remaining vector interaction contributes to the spin independent scattering cross section. The couplings $g_{\eta^{0R}\eta^{0I}Z^0}=ig/(2c_W)$ and $g_{\bar{q}qZ^0}=g(I_q-2e_qs_W^2)/(2c_W)$ are fixed by the SU(2)$_L\times$U(1)$_Y$ gauge couplings $g$ and $g'$, their ratio $\tan\theta_W=s_W/c_W=g'/g$ and the quark weak isospin $I_q=\pm1/2$ and fractional charge $e_{u,d}=(2/3,-1/3)$ and can be obtained using \textsc{SARAH}  \cite{Staub:2013tta}.

Integrating out the mediator produces the effective Lagrangian \cite{Agrawal:2010fh} 
\begin{align}
    \mathcal{L}_\text{eff}=-b_q(\eta^{0R}\partial_\mu\eta^{0I}-\partial_\mu\eta^{0R}\eta^{0I})\Bar{q}\gamma^\mu q,
\end{align}
where $b_q:=g_{\eta^{0R}\eta^{0I}Z^0}g_{\bar{q}qZ^0}/m_Z^2$. In this case, the effective operator already has the correct mass dimension. The scattering amplitude is then 
\begin{align}
  \mathcal{M}=4m_{\eta^{0R}}m_N\sum_qb_q\delta^0_\mu\langle N|\bar{q}\gamma^\mu q|N\rangle,
\end{align}
since the scalar four-momenta add up to $p^{\eta^{0R}}_\mu+p^{\eta^{0I}}_\mu\approx 2m_{\eta^{0R,I}}\delta_\mu^0$. Due to the conservation of the vector current, only valence quark contributions must be considered.

Using the non relativistic normalization for nucleon states, the vector matrix element for a nucleon with $n_q$ valence ($u,d$) quarks is simply \cite{Agrawal:2010fh} 
\begin{align}
  2m_N\langle N|\bar{q}\gamma^\mu q|N\rangle=n_q\bar{u}_N\gamma^\mu u_N\approx n_q 2m_N\delta^\mu_0.
\end{align}
The valence quark contributions thus add up coherently to 
\begin{align}
    b_p&=2b_u+b_d,\\
    b_n&=b_u+2b_d.
\end{align}
Note that for vector interactions, in contrast to the Higgs exchange, the scattering off protons and neutrons differs significantly.

Since the vector interaction is spin independent, we can also sum coherently over the contributions of all nucleons in the nucleus. The cross section then takes a similar form as for the scalar interaction, i.e.\ \cite{Agrawal:2010fh} 
\begin{align}
    \sigma_A^0=\frac{\mu^2}{\pi}\left[Zb_p+(A-Z)b_n\right]^2.
\end{align}
Since the inelastic scattering is mediated by a $Z^0$-boson and the corresponding couplings are SM gauge couplings, the resulting cross sections are significantly larger than the elastic Higgs-mediated cross sections. 

While the inelasticity can be safely neglected in the differential cross section
\begin{align}
  \frac{d\sigma_A}{d|{\bf q}|^2}=\frac{\sigma_A^0}{4\mu^2v^2}F^2(Q),
\end{align}
and form factor, it does affect the kinematics through the integration boundaries \cite{Nussinov:2009ft,Menon:2009qj,Shu:2010ta} 
\begin{align}
	Q_\text{min}&=\frac{1}{2}m_{\eta^{0R,I}} v^2\left(1-\frac{\mu^2}{m_A^2}\left(1+\frac{m_A}{m_{\eta^{0R,I}}}\sqrt{1-\frac{\delta}{\mu v^2/2}}\right)^2\right)-\delta,\label{eq:Q_min}\\
  Q_\text{max}&=\frac{1}{2}m_{\eta^{0R,I}} v^2\left(1-\frac{\mu^2}{m_A^2}\left(1-\frac{m_A}{m_{\eta^{0R,I}}}\sqrt{1-\frac{\delta}{\mu v^2/2}}\right)^2\right)-\delta.\label{eq:Q_max}
\end{align}

\subsection{Capture rate}

The DM capture rate per unit shell volume in a celestial body such as the Sun is given by \cite{Gould:1987,Nussinov:2009ft,Menon:2009qj,Shu:2010ta} 
\begin{align}
  \frac{dC}{dV}=\int_0^\infty\ du \frac{f(u)}{u}w\Omega_v^-(w),
\end{align}
where $f(u)$ is the DM velocity distribution outside of the gravitational field, which we assume to follow the Maxwell-Boltzmann distribution, and $u$ is the DM velocity at infinity. After falling into the gravitational potential of the Sun, the DM velocity is 
\begin{align}
  w=\sqrt{u^2+v(r)^2},
\end{align}
where $v(r)$ is the escape velocity at a shell with radius $r$. The rate per unit time, with which a DM particle with velocity $w$ scatters off a nucleus $A$ to a velocity less than $v(r)$, is 
\begin{align}
  \Omega_v^-=\frac{n_A \sigma_A^0 w}{4\mu^2v^2/(2m_A)} \int_{Q'_{\min}}^{Q_{\max}}\ dQ F^2(Q).\label{eq. Scattfkt inel}
\end{align} 
It is determined by the total scattering rate $n_A \sigma_A^0 w$, where $n_A$ denotes the number density of the nucleus $A$ in the Sun, and a conditional probability, that ensures the capture. For inelastic scattering, non relativistic kinematics constrain the energy transfer $Q$ to lie within the limits of Eqs.\ (\ref{eq:Q_min}) and (\ref{eq:Q_max}), where $v$ should be replaced by $w$. For the DM particle to be captured, the energy transfer must in addition be greater than 
\begin{align}
  Q_\text{cap}&=\frac{1}{2}m_{\eta^{0R,I}} (w^2-v(r)^2)-\delta.
\end{align}
so that $Q'_{\min}=\max(Q_{\rm cap},Q_{\min})$. The well-known results for elastic scattering are recovered in the limit $\delta\to0$ \cite{Gould:1987}. For small mass splittings below $\mathcal{O}(100 \text{ keV})$ the capture rate is almost constant. For larger mass splittings, the capture rate quickly drops off, as the inelastic scattering becomes kinematically forbidden.\footnote{We have reproduced the numerical results for the capture rate in Fig.\ 3 of Ref.\ \cite{Menon:2009qj} up to a missing factor of two in $E_{\max}^{\rm elastic}$ as well as those in Figs.\ 1 and 2 of Ref.\ \cite{Shu:2010ta}. Note that Ref.\ \cite{Nussinov:2009ft} uses $Q'_\text{min}=Q_\text{cap}$, which ignores the case where $Q_{\min}>Q_{\rm cap}$ and thus gives too large capture rates for large mass splittings.}

The time evolution of the DM population $N$ in the Sun is controled by
\begin{equation}
    \Dot{N} = C - 2 \Gamma,
\end{equation}
where the capture rate $C$ depends on the elastic and inelastic DM-nucleus scattering cross sections and nuclear abundances as described above and $\Gamma$ is the annihilation rate. The evaporation rate relevant for very light DM has been neglected. Over time, DM capture and annihilation will reach equilibrium. Provided that the DM-nucleus scattering cross section is sufficiently large, we can assume that the time of equilibrium onset is well below the age of the Sun \cite{Blennow:2018xwu}. This leads to an annihilation rate of
\begin{equation}
  \Gamma = C /2.
\end{equation}
Hence the DM annihilation rate in the Sun depends directly on the elastic and inelastic scattering cross sections and on the nuclear abundances in the Sun.

\section{Indirect detection of elastic and inelastic DM in the Sun with \textsc{IceCube}\label{sec: Detection IceCube}}

Our numerical analysis of elastic and inelastic DM and in particular of the expected neutrino signals at \textsc{IceCube} from DM annihilations in the Sun is based on an implementation of the model described in Sec.\ \ref{sec: The model} in \textsc{Sarah 4.14.0} \cite{Staub:2013tta}. The physical mass spectrum and branching ratios, in particular those for LFV processes, are computed with \textsc{SPheno 4.0.3} \cite{Poro03,PS11}. The DM relic density, direct detection cross sections and neutrino event rates are obtained from \textsc{micrOMEGAs 5.0.8} \cite{BBGPZ18}.

Assuming equilibrium of capture and annihilation in the core of the Sun ($\Gamma=C/2$), the differential flux of neutrinos or antineutrinos on Earth is given by \cite{JUNGMAN1996195,Belanger:2015hra} 
\begin{equation}
\frac{d\phi_{\nu}}{dE_{\nu}}=\frac{1}{4\pi d_{\odot}^{2}}\Gamma\sum_{f}Br_{f\bar{f}}\frac{dN_{f}}{dE_{\nu}},\label{eq: flux sun earth with branching}
\end{equation}
where $d_{\odot}$ is the distance Earth-Sun, $Br_{f\bar{f}}$ are the branching fractions into particle-antipar\-ticle final states $f\bar{f}$, and $dN_{f}/dE_{\nu}$ are the corresponding neutrino (antineutrino) energy spectra. In \textsc{micrOMEGAs}, the latter are computed based on tables and feature neutrino propagation and oscillation in the Sun and in vacuum. The function \texttt{neutrinoFlux} automatizes the process of calculating the capture rate, the annihilation branching ratios and spectra of the respective channels and provides the total neutrino flux at Earth. Since the inelastic scenario was not implemented in \textsc{micrOMEGAs}, we used \textsc{CalcHEP 3.7} \cite{Belyaev:2012qa} to compute the corresponding DM-quark scattering matrix elements and cross sections as described in the previous section. We then modified the routine \texttt{dssenu\underline{ }capsunnum} in \textsc{DarkSUSY 6.2.3} \cite{Bringmann:2018lay} to obtain the inelastic capture rate, which was then fed back to \textsc{micrOMEGAs}.

\begin{figure}
  \centering
  \includegraphics[width=0.63\textwidth]{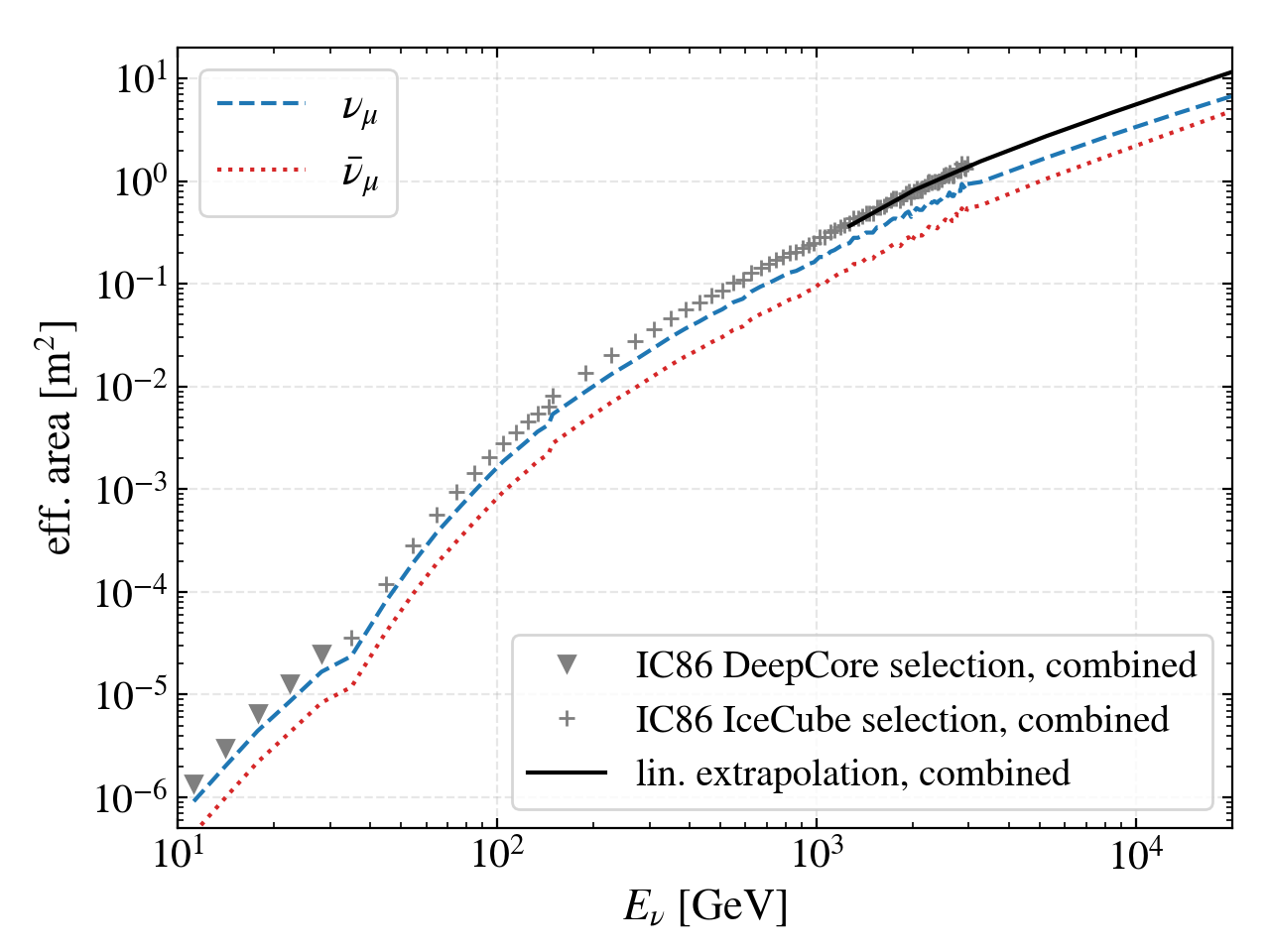}
  \caption{\label{plot: IC86 Aeff} The $\nu_{\mu}$ and $\bar{\nu}_{\mu}$ effective areas of the \textsc{DeepCore} detector and the \textsc{IceCube} detector configuration \textsc{ic86} as a function of the neutrino energy. The data for the combined effective area is taken from Ref.\ \cite{Aartsen:2016zhm} (triangles and crosses) and linearly extrapolated (solid black line). The individual effective areas for neutrinos (dashed blue line) and antineutrinos (dotted red line) are calculated with the deep inelastic scattering cross sections taken from Ref.\ \cite{Buga2003}. Both the \textsc{IceCube} and \textsc{DeepCore} selections of the effective area are used in our work.}
\end{figure}

The differential number of signal events in the detector is given by \cite{Belanger:2015hra} 
\begin{equation}
\frac{dN_{s}}{dE}=t_{e}\left(\frac{d\phi_{\nu_{\mu}}}{dE}A_{\nu_{\mu}}(E)+\frac{d\phi_{\bar{\nu}_{\mu}}}{dE}A_{\bar{\nu}_{\mu}}(E)\right)\,,\label{eq:Nsig}
\end{equation}
where $t_{e}$ is the exposure time and $A_{\nu_{\mu}(\bar{\nu}_{\mu})}$ is the muon neutrino (muon antineutrino) effective area of the detector. A routine for the effective area of the now obsolete configuration \textsc{ic22} (where 22 is the number of data-taking strings), \texttt{IC22nuAr}, was already implemented in \textsc{micrOMEGAs}. We updated this routine using the data from Ref.\ \cite{Aartsen:2016zhm} for the effective area of \textsc{ic86}. Eight DeepCore strings are part of the \textsc{IC86} configuration. Including their effective area lowers the energy threshold to 10 GeV. We extrapolated the data points linearly to fit our energy range, as shown in Fig.\ \ref{plot: IC86 Aeff}.
The corresponding data points have been taken from Ref.\ \cite{Aartsen:2016zhm}. In the region where both selections overlap, we use the \textsc{IceCube} effective area, as it is larger than the one of the \textsc{DeepCore} selection \cite{deBoer:2021xjs}.

\textsc{IceCube} is sensitive to both $\nu_{\mu}$ and $\bar{\nu}_{\mu}$. However, the corresponding energy dependent deep-inelastic scattering cross sections with nucleons in the detector are slightly different. Since Ref.\ \cite{Aartsen:2016zhm} provides only the combined $\nu_{\mu}+\bar{\nu}_{\mu}$ effective area, we calculate the individual effective areas by taking into account the different cross sections $\sigma_{\nu_{\mu}\left(\bar{\nu}_{\mu}\right)}$ given in Ref. \cite{Buga2003} with the relation
\begin{equation}
A_{\nu_{\mu}\left(\bar{\nu}_{\mu}\right)}=\frac{A_{\text{combined}}}{1+\frac{\sigma_{\bar{\nu}_{\mu}\left(\nu_{\mu}\right)}}{\sigma_{\nu_{\mu}\left(\bar{\nu}_{\mu}\right)}}}\,.\label{eq: Aeff}
\end{equation}

We illustrate the expected neutrino fluxes at Earth and event rates in \textsc{IceCube} for a specific benchmark point BPA in the scotogenic model. The corresponding parameters are listed in Tab.\ \ref{tab: BPA parameters}.
\begin{table}
\begin{centering}
\caption{\label{tab: BPA parameters}Parameters of the scotogenic model for our benchmark point BPA. Shown are the coupling parameters $\lambda_i$, the squared mass $m_{\eta}^{2}$ of the new scalar doublet, the (diagonal) mass matrix $m_{N}$ for the three new fermion singlets, and the real and imaginary parts of the Yukawa matrix, $y^{R}$ and $y^{I}$.}
\renewcommand{\arraystretch}{1.5}
{\footnotesize{}}%
\begin{tabular}{|ccccc|c|ccc|}
\hline 
{\footnotesize{}$\lambda_{1}$} & {\footnotesize{}$\lambda_{2}$} & {\footnotesize{}$\lambda_{3}$} & {\footnotesize{}$\lambda_{4}$} & {\footnotesize{}$\lambda_{5}$} & {\footnotesize{}$m_{\eta}^{2}$} & {\footnotesize{}$m_{N_{1}}$} & {\footnotesize{}$m_{N_{2}}$} & {\footnotesize{}$m_{N_{3}}$}\tabularnewline
\hline 
{\footnotesize{}$0.26$} & {\footnotesize{}$0.50$} & {\footnotesize{}$0.56$} & {\footnotesize{}$-0.14$} & {\footnotesize{}$2.00\cdot10^{-7}$} & {\footnotesize{}$1.00\cdot10^{6}$} & {\footnotesize{}$1.32\cdot10^{3}$} & {\footnotesize{}$3.13\cdot10^{3}$} & {\footnotesize{}$3.44\cdot10^{3}$}\tabularnewline
\hline 
\hline 
\multicolumn{5}{|c|}{{\footnotesize{}$y^R/10^{-2}$}} & \multicolumn{4}{c|}{{\footnotesize{}$y^I/10^{-3}$}}\tabularnewline
\hline 
\multicolumn{5}{|c|}{{\footnotesize{}$\left(\begin{array}{ccc}
-17.20 & 2.07 & -6.91\\
-4.94 & 9.41 & 15.17\\
5.22 & 15.98 & -8.20
\end{array}\right)$}} & \multicolumn{4}{c|}{{\footnotesize{}$\left(\begin{array}{ccc}
2.58 & 4.46 & 5.10\\
-6.88 & 2.02 & 2.30\\
-2.08 & -1.23 & -1.40
\end{array}\right)$}}\tabularnewline
\hline 
\end{tabular}{\footnotesize\par}
\end{centering}
\end{table}
With these parameters, we obtain the correct relic density of $\Omega h^{2}=0.1217$ for the scalar DM candidate $\eta^{0I}$ with mass 1007.38 GeV. In descending order of importance, the DM annihilation channels are $W^{+}W^{-}$ (58.36\%), $Z^{0}Z^{0}$ (23.51\%), $hh$ (12.30\%) and $W^{+}W^{-}\gamma$ (4.99\%), i.e.\ $W^{+}W^{-}$ pairs (plus an accompanying photon) represent the dominant channel with a branching fraction of over 63\%, as is the case for most points in the parameter space with $m_{\eta^{0R,I}}>m_W$. Direct annihilation into neutrinos is suppressed by both the small Yukawa couplings $y_{i\alpha}$ and the large neutrino propagator masses $m_{N_i}$.

\begin{figure}
  \centering
  \includegraphics[width=0.63\textwidth]{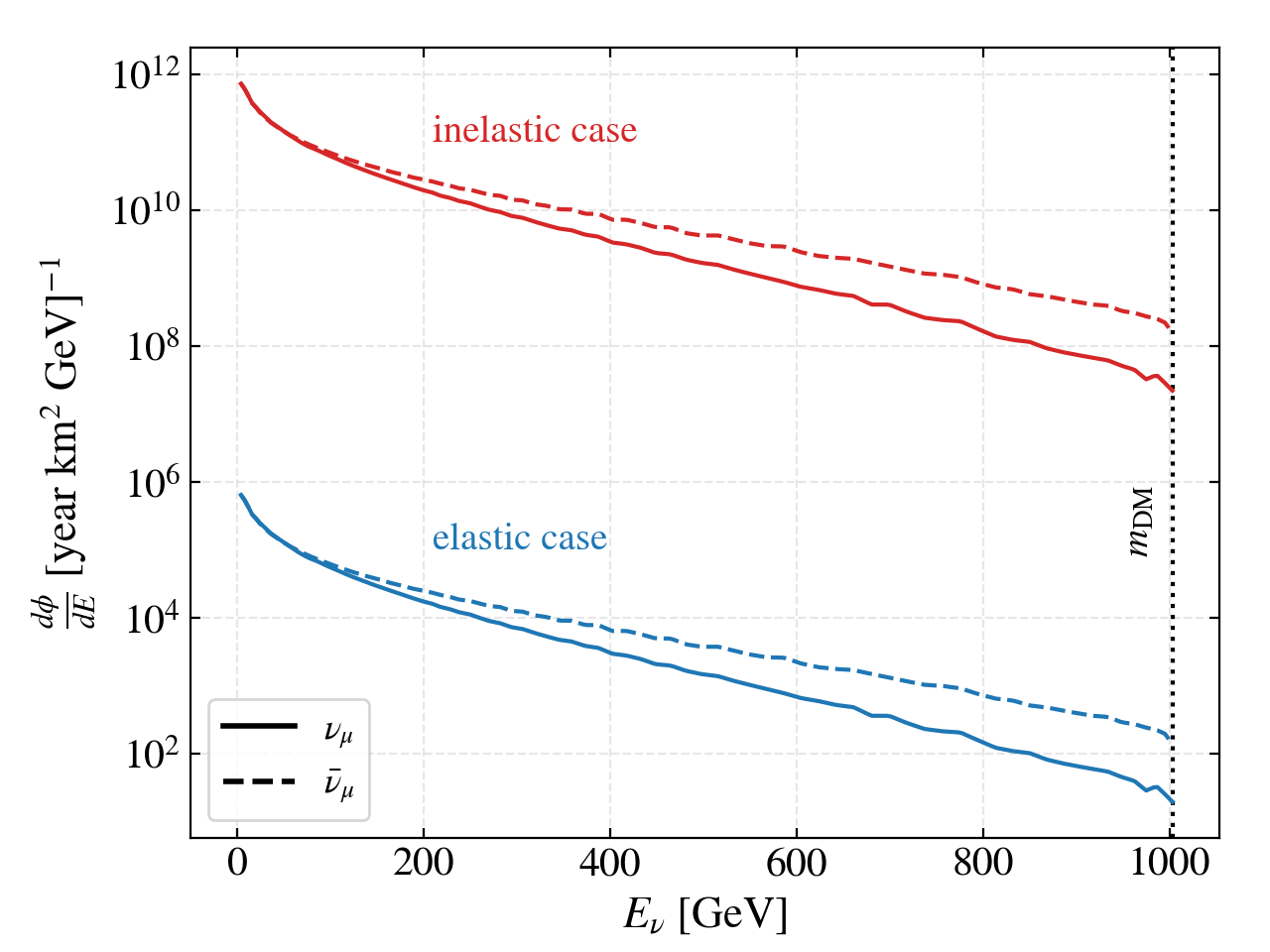}
  \includegraphics[width=0.63\textwidth]{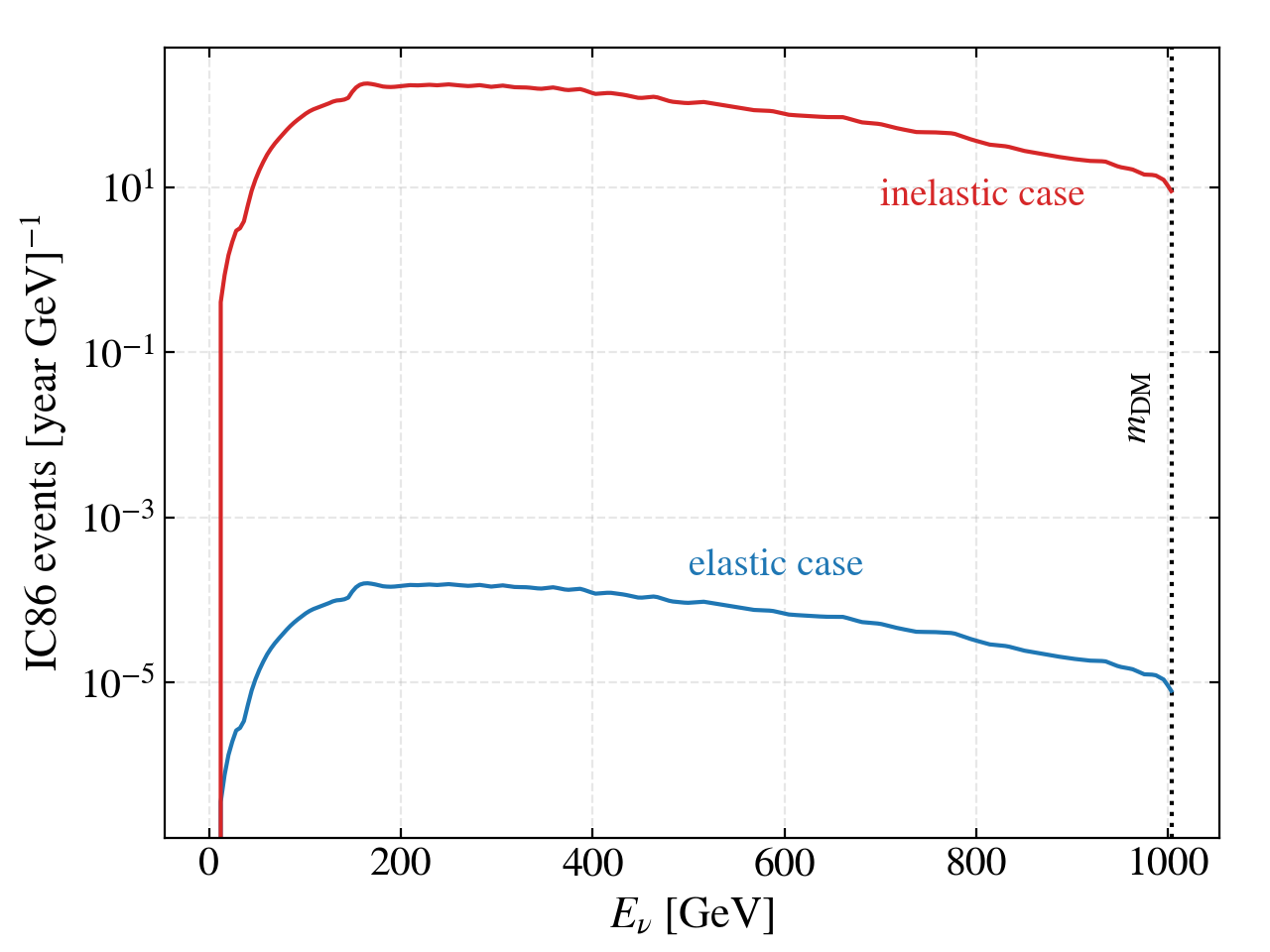}
  \caption{\label{plot: BSA spectrum flux}Top: Elastic (blue) and inelastic (red) muon neutrino (full) and antineutrino (dashed) fluxes at Earth as a function of the neutrino energy, calculated with the modified function \texttt{neutrinoFlux} in \textsc{micrOMEGAs} at the parameter point BPA. The mass of the DM particle $\eta^{0I}$ is 1007.38 GeV. Bottom: Expected number of signal events in \textsc{ic86} per year as a function of the neutrino energy.}
\end{figure}

We show the differential neutrino and antineutrino fluxes at BPA for both elastic (blue) and inelastic (red) scattering in the Sun in Fig.\ \ref{plot: BSA spectrum flux} (top).
The differences in differential fluxes between neutrinos (full) and antineutrinos (dashed lines) that show up at high energies are due to absorption (also oscillation and regeneration) effects taking place inside the Sun \cite{Cirelli:2005gh}. Because there is no annihilation into neutrinos directly, we do not observe a monochromatic neutrino line in the spectrum, but instead a sharp cut-off at the DM mass of about 1 TeV. The differential number of expected signal events per year in \textsc{ic86} is shown in Fig.\ \ref{plot: BSA spectrum flux} (bottom). After integration over the neutrino energy, BPA yields a total number of only $0.07$ expected events per year in \textsc{ic86} in the elastic case, but $8.65\cdot10^{4}$ in the inelastic case.

\section{Numerical scan\label{sec: Parameter scans}}

Using the tool chain described in the previous chapter, we now perform a numerical scan of the scotogenic parameter space. To comply with the requirements of vacuum stability, perturbativity and that the lightest scalar must be neutral, the scalar couplings are varied within the ranges 
\begin{equation}
  \lambda_{3}\in\left[-\sqrt{\lambda_{1}\lambda_{2}},4\pi\right]\ ,\
  \lambda_{4} \in \left[\max\{-\sqrt{\lambda_{1}\lambda_{2}}-\lambda_{3}+\left|\lambda_{5}\right|,-4\pi\},0\right] \ ,\
  \left|\lambda_{5}\right|\in\left[10^{-10},10^{-3}\right]
\end{equation}
As was mentioned in Sec.\ \ref{sec: The model}, $\lambda_{5}$ is naturally small, since the SM neutrino masses would vanish and the symmetry of the Lagrangian increase if it was exactly zero. The mass parameter of the new scalar doublet is varied in the range
\begin{equation}
  m_{\eta}\in\left[\unit[1]{GeV},\unit[10]{TeV}\right].
\end{equation}
The masses $m_{N_{i}}$ of the new fermion singlets are required to be larger than the mass of the lightest scalar, but below $10$ TeV. In a second scan, we choose this mass difference to be small (i.e.\ 0.1 GeV) in order to demonstrate the impact of coannihilation processes \cite{KRRYZ13}. The SM neutrino oscillation parameters are taken from Ref.\ \cite{Esteban:2018azc} in the $3\sigma$ range, assuming normal ordering. The CP violating phase $\delta_{CP}$ is varied from zero to $2\pi$, and the lightest neutrino mass is varied in the interval $[10^{-4},1.1]$ eV in accordance with the most recent limit set by \textsc{KATRIN} \cite{Aker:2019uuj}. The Yukawa couplings are then calculated using the Casas-Ibarra parametrization and required to satisfy $|y_{i\alpha}|^2<4\pi$.

The scotogenic model is further constrained by a number of other experimental measurements. In particular, we impose the relic density measurement by \textsc{Planck} \cite{Agha18} with a relatively loose margin of $\Omega h^{2}=0.12\pm0.02$ in order to account for theoretical uncertainties \cite{Harz:2016dql}. We also impose the LFV branching ratios (BR) and conversion rate (CR)
of 
\begin{eqnarray}
\text{BR}\left(\mu\rightarrow e+\gamma\right) & < & 4.2\cdot10^{-13},\nonumber \\
\text{BR}\left(\mu\rightarrow3e\right) & < & 1.0\cdot10^{-12},\label{eq: LFV constraints}\\
\text{CR}\left(\mu-e,\text{Ti}\right) & < & 4.3\cdot10^{-12},\nonumber 
\end{eqnarray}
published by the MEG \cite{Baldi16}, \textsc{SINDRUM} \cite{Bell88} and \textsc{SINDRUM II} \cite{Dohm93} collaborations. Furthermore, we apply limits on the new physics invisible decay width of the $Z^0$ boson from LEP \cite{Carena:2003aj}
\begin{equation}
  {\rm BR}(Z^0\to{\rm new})< 0.008,
\end{equation}
which effectively excludes $m_{\rm DM}< m_Z/2$ \cite{Cao:2007rm,Lundstrom:2008ai}, and on the invisible decay width of the SM Higgs boson from ATLAS (CMS) at the LHC \cite{ATLAS:2020kdi,Sirunyan:2018owy}
\begin{equation}
  {\rm BR}(h\to{\rm inv.})< 0.11\ (0.19),
\end{equation}
as well as on the elastic and inelastic scattering cross section from direct searches with \textsc{XENON1T} \cite{Aprile:2018dbl}, \textsc{XENON100} \cite{Aprile:2011ts,Aprile:2017aas} and \textsc{PandaX-II} \cite{Chen:2017cqc}. The possible signal from \textsc{DAMA/LIBRA} \cite{Bernabei:2018jrt,Bernabei:2019ajy}, its ongoing verification \cite{deSouza:2016fxg,Adhikari:2019off,Amare:2021yyu}, previous limits on indirect detection from DM annihilation into neutrinos in the Sun \cite{Adrian-Martinez:2016gti,Aartsen:2016zhm,Choi:2015ara} and the Galactic Center \cite{Albert:2016emp,Aartsen:2017ulx,Aartsen:2020tdl,Abe:2020sbr}, and expected event rates for neutrinos from the Sun in the current \textsc{IceCube} configuration with 86 strings (\text{IC86}) are also discussed in the following.

\subsection{Limits on the elastic cross section}

\begin{figure}
 \centering
 \includegraphics[width=0.6\textwidth]{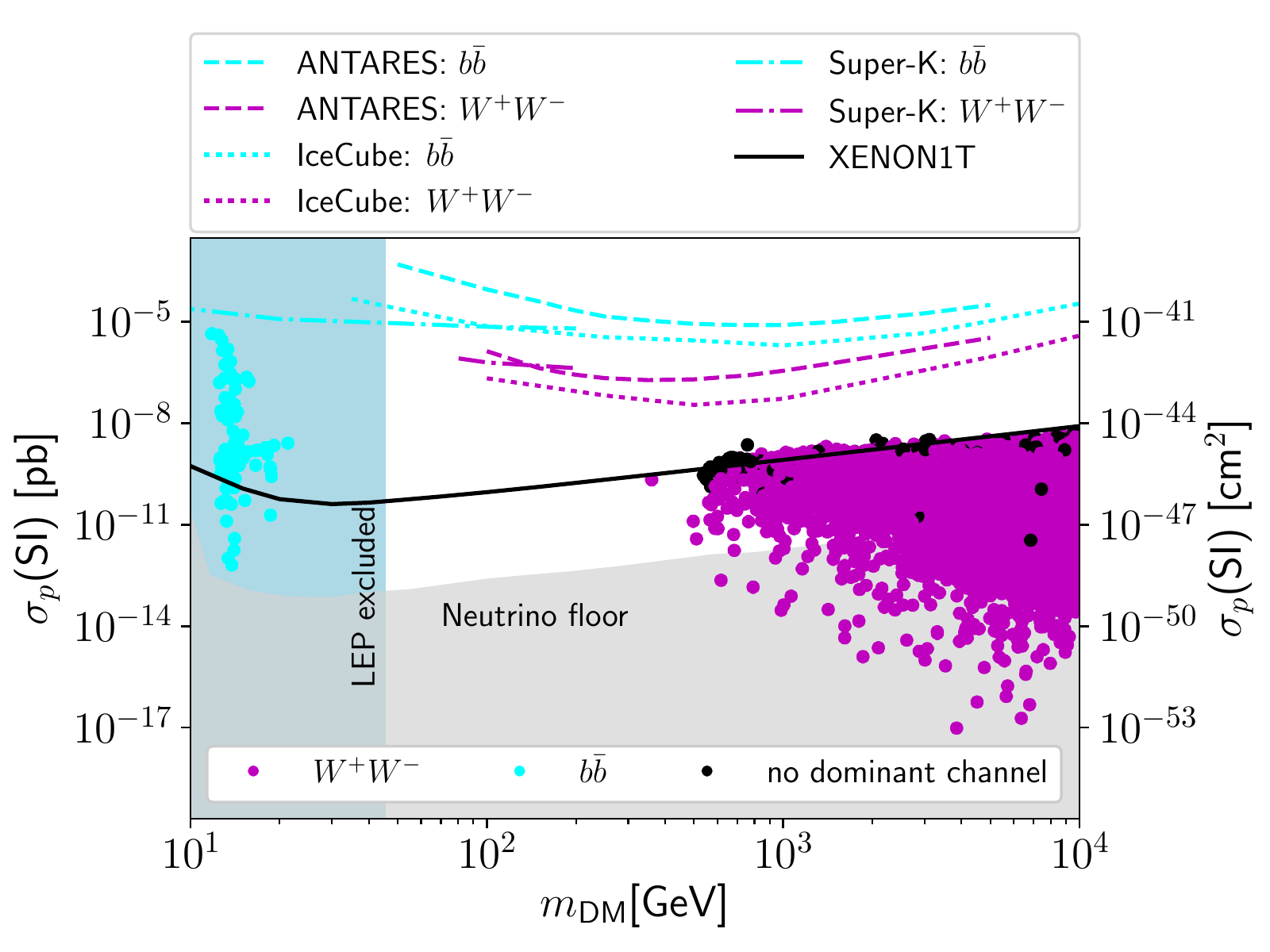}
 \includegraphics[width=0.6\textwidth]{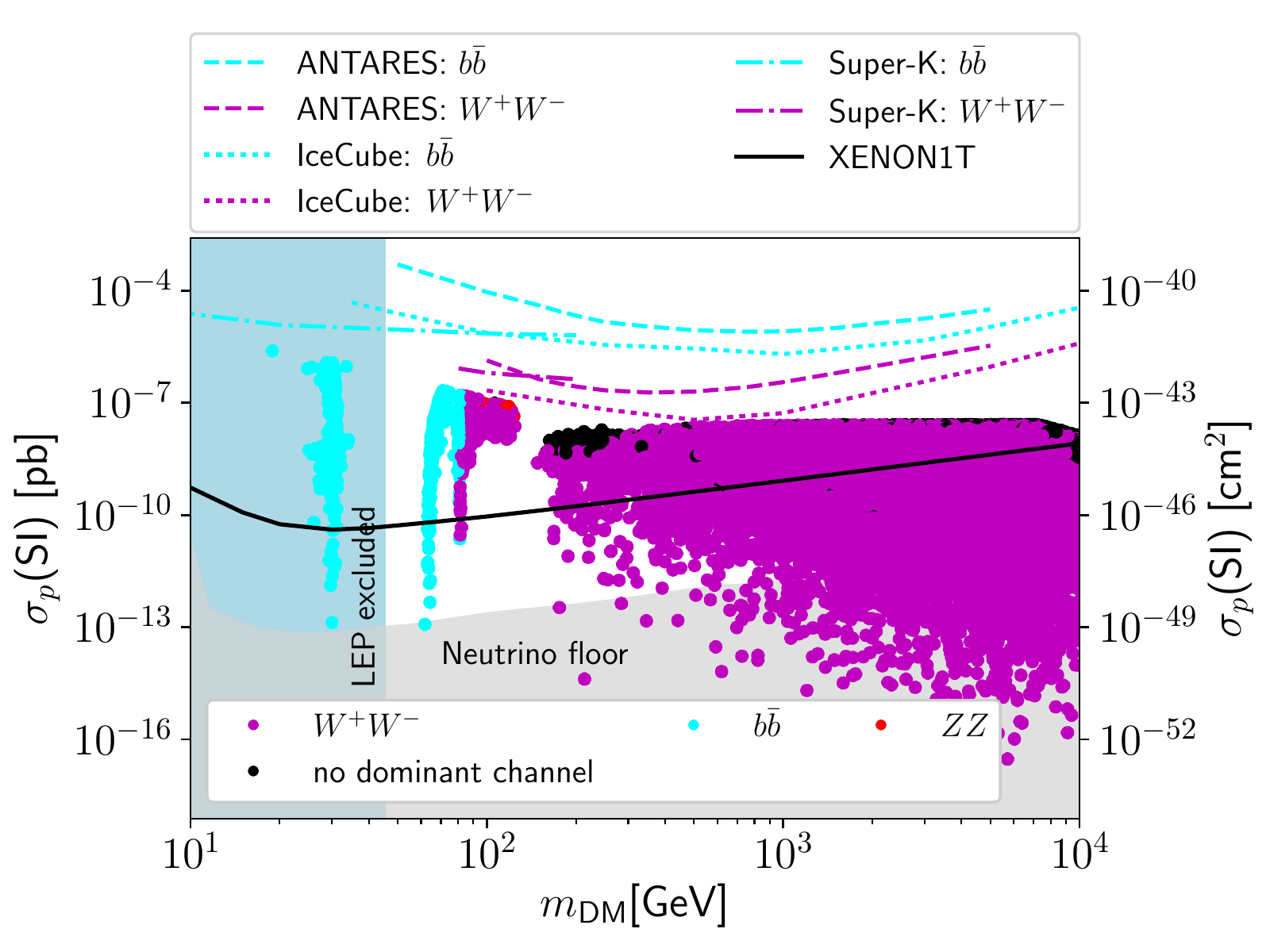}
 \caption{The spin independent (SI) elastic cross section with \textsc{ANTARES} \cite{Adrian-Martinez:2016gti}, \textsc{IceCube} \cite{Aartsen:2016zhm}, \textsc{Super-Kamiokande} \cite{Choi:2015ara} and \textsc{XENON1T} \cite{Aprile:2018dbl} exclusion limits as a function of the DM mass. All points and lines are color coded according to the main annihilation channel, provided there is one with a branching ratio of over 50\%. Also shown are the LEP exclusion from the invisible $Z^0$ boson width \cite{Cao:2007rm,Lundstrom:2008ai} and the neutrino floor \cite{Billard:2013qya}. In the lower plot, coannihilation processes are enhanced by the small scalar-fermion mass difference.}
 \label{fig:5}
\end{figure}

Our spin independent elastic cross section results for scalar DM in the scotogenic model are shown in Fig.\ \ref{fig:5}. Apart from the larger mass range of our scan and considering that we imposed updated constraints from the Higgs mass, DM relic density, neutrino masses/mixings and LFV, our results agree with those in Fig.\ 9 in Ref.\ \cite{KRRYZ13}. In particular, models with low mass DM contain sufficiently heavy charged scalars to evade the corresponding LEP2 bound \cite{Abbi03}, but are excluded by the LEP limit on the invisible $Z^0$ decay width \cite{Cao:2007rm,Lundstrom:2008ai}. Since there are no viable models with $m_Z/2<m_{\rm DM}<m_h/2$, there are no additional constraints from the invisible $h$ decay width \cite{ATLAS:2020kdi,Sirunyan:2018owy}. Otherwise and as well known, the correct relic density requires the DM to be larger than about 500 GeV (top). Fermion coannihilation processes reduce this lower mass limit to about 200 GeV (bottom) \cite{KRRYZ13}. The sampled points are color coded according to the dominant annihilation branching ratio. A point is marked as having no dominant channel when no single branching ratio reaches 50\%. Above the $W$ boson threshold, mostly annihilation into $W$ boson pairs occurs. The channel-dependent constraints derived from indirect DM detection with \textsc{ANTARES} \cite{Adrian-Martinez:2016gti}, \textsc{IceCube} \cite{Aartsen:2016zhm} and \textsc{Super-Kamiokande} \cite{Choi:2015ara} are considerably weaker than the direct detection constraints from \textsc{XENON1T} \cite{Aprile:2018dbl}. Only the latter constrain the parameter space, and in particular the coannihilation region. Since the Higgs coupling to the quarks in the nucleon is relatively small, the elastic cross sections are small as well. A significant part of the parameter space for DM masses beyond 1 TeV results even in cross sections below the atmospheric and diffuse supernova background (DSNB) neutrino ``floor'' \cite{Billard:2013qya}, which may render DM direct detection difficult.

\subsection{Limits on the inelastic cross section}

\begin{figure}
 \centering
 \includegraphics[width=0.6\textwidth]{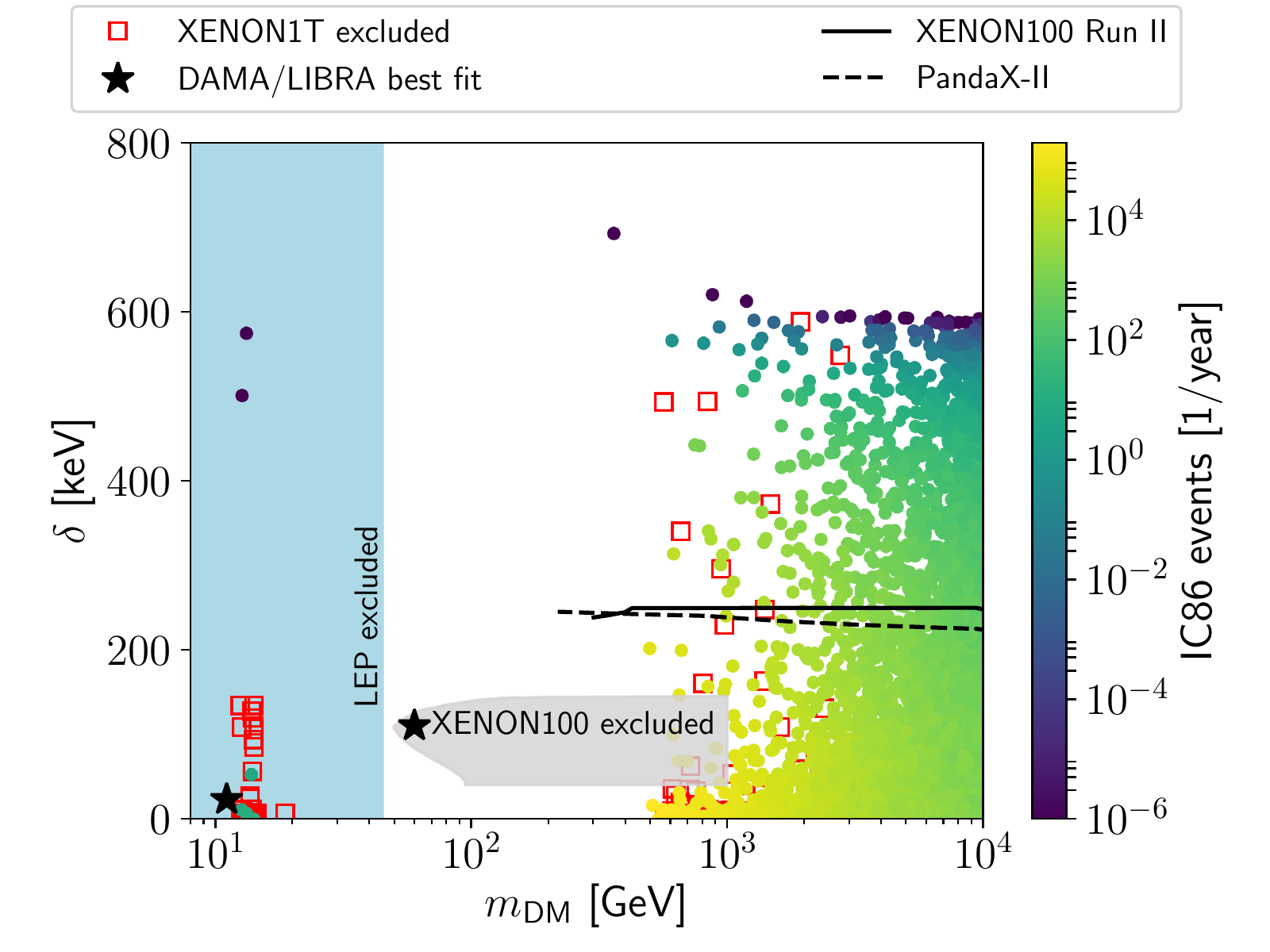}
 \includegraphics[width=0.6\textwidth]{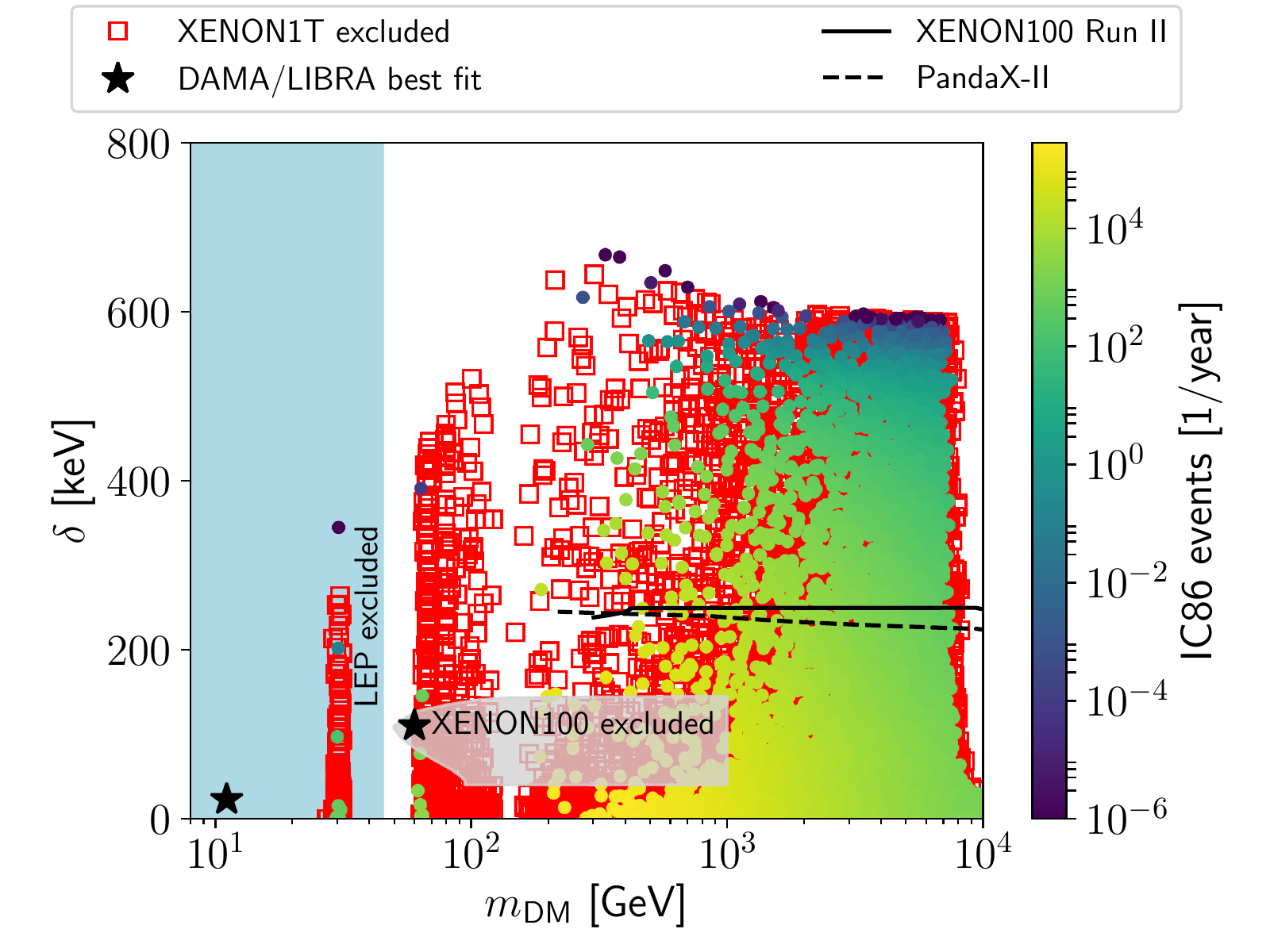}
 \caption{Scotogenic models in the plane DM mass vs.\ neutral scalar mass splitting, color coded for expected IC86 events (temperature scale) and elastic scattering exclusion by \textsc{XENON1T} (red boxes) \cite{Aprile:2018dbl}. Also shown are the exclusion of the low- (sodium) \cite{Kang:2019uuj} and high-mass (iodine) \textsc{DAMA/LIBRA} preferred regions \cite{Bernabei:2018jrt,Bernabei:2019ajy} by \textsc{LEP} and \textsc{XENON100} \cite{Aprile:2011ts} and the upper limits on the mass splitting from the \textsc{XENON100} Run II data \cite{Aprile:2017aas} and by the \textsc{PandaX-II} experiment with smaller recoil energy window and larger background \cite{Chen:2017cqc}. In the lower plot, coannihilation processes are enhanced by the small scalar-fermion mass difference.}
 \label{fig:6}
\end{figure}

The observation of an annual modulation signal by \textsc{DAMA/LIBRA} \cite{Bernabei:2018jrt,Bernabei:2019ajy}, its interpretation as DM of low or (relatively) higher mass of about 10 and 50 GeV \cite{Kahlhoefer:2018knc,Kang:2018qvz} and the tension with other direct detection experiments have led to the speculation that DM might only undergo inelastic scattering off nuclei in \textsc{DAMA/LIBRA}, to which other experiments would not be sensitive \cite{TuckerSmith:2001hy}. To fit the \textsc{DAMA/LIBRA} observation, the inelastic nucleon cross section was usually assumed to be 10$^{-4}$ pb \cite{Nussinov:2009ft,Menon:2009qj,Shu:2010ta}. This value is indeed close to the proton cross section of $1.7\cdot 10^{-4}$ pb in the scotogenic model, where the mediator (as shown in Fig. \ref{diagram: scotogenic SISD scattering}) is not a Higgs ($h$) boson, but an electroweak gauge ($Z^0$), which violates isospin by $b_n/b_p=1/(4\sin^2\theta_W-1)\simeq-6.6$.\footnote{The neutron cross section $1/(2\pi)G_F^2m_N^2 = 74.3\cdot 10^{-4}$ pb is therefore significantly larger \cite{Arina:2009um,Hashimoto:2020xoz}.} It also corresponds to typical cross sections in the inert doublet model, which is in fact the only model with a {\em single} scalar DM multiplet that allows for naturally small neutral component mass splittings at the renormalizable level \cite{Arina:2009um}.

\begin{figure}
 \centering
 \includegraphics[width=0.6\textwidth]{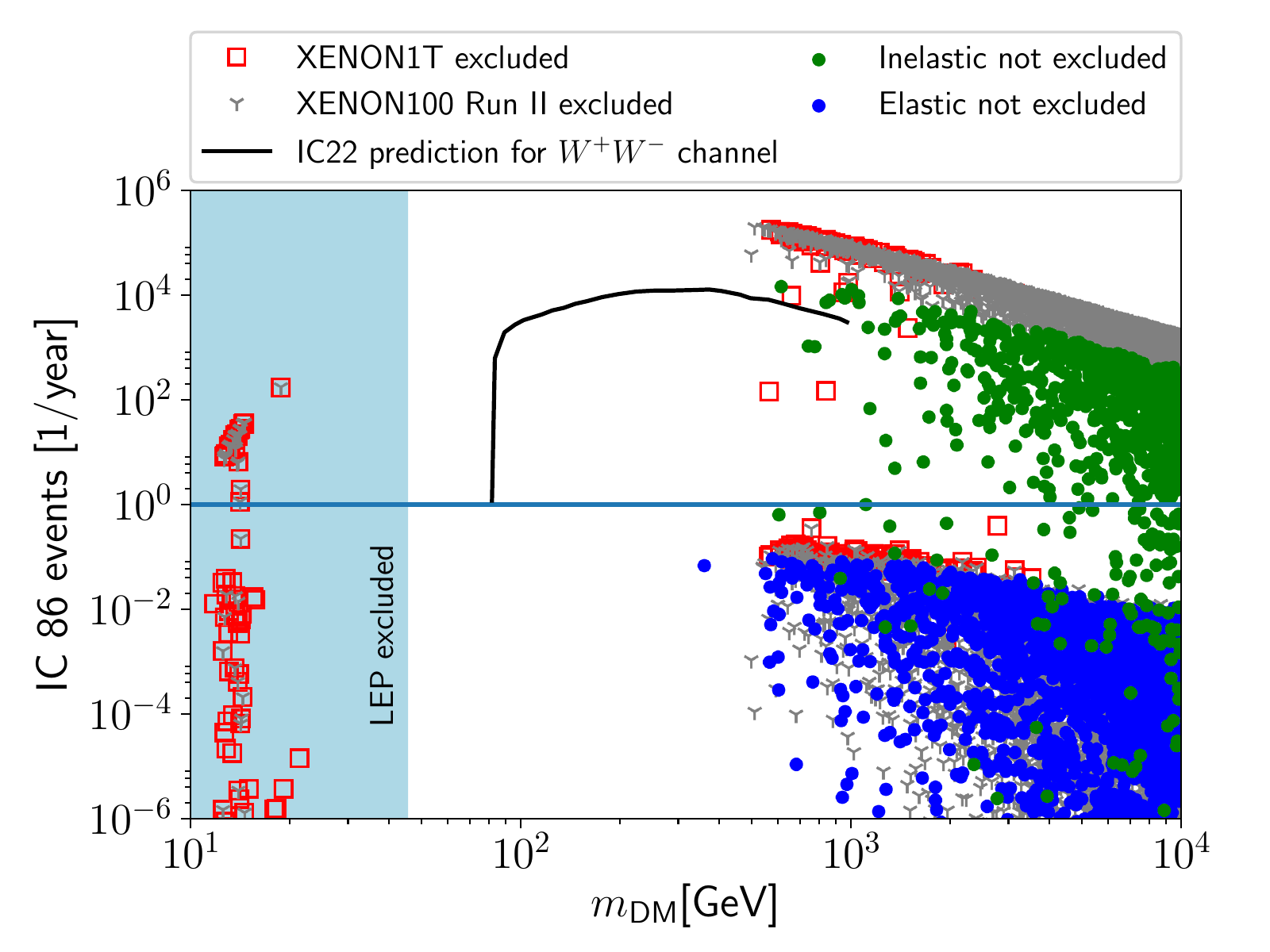}
 \includegraphics[width=0.6\textwidth]{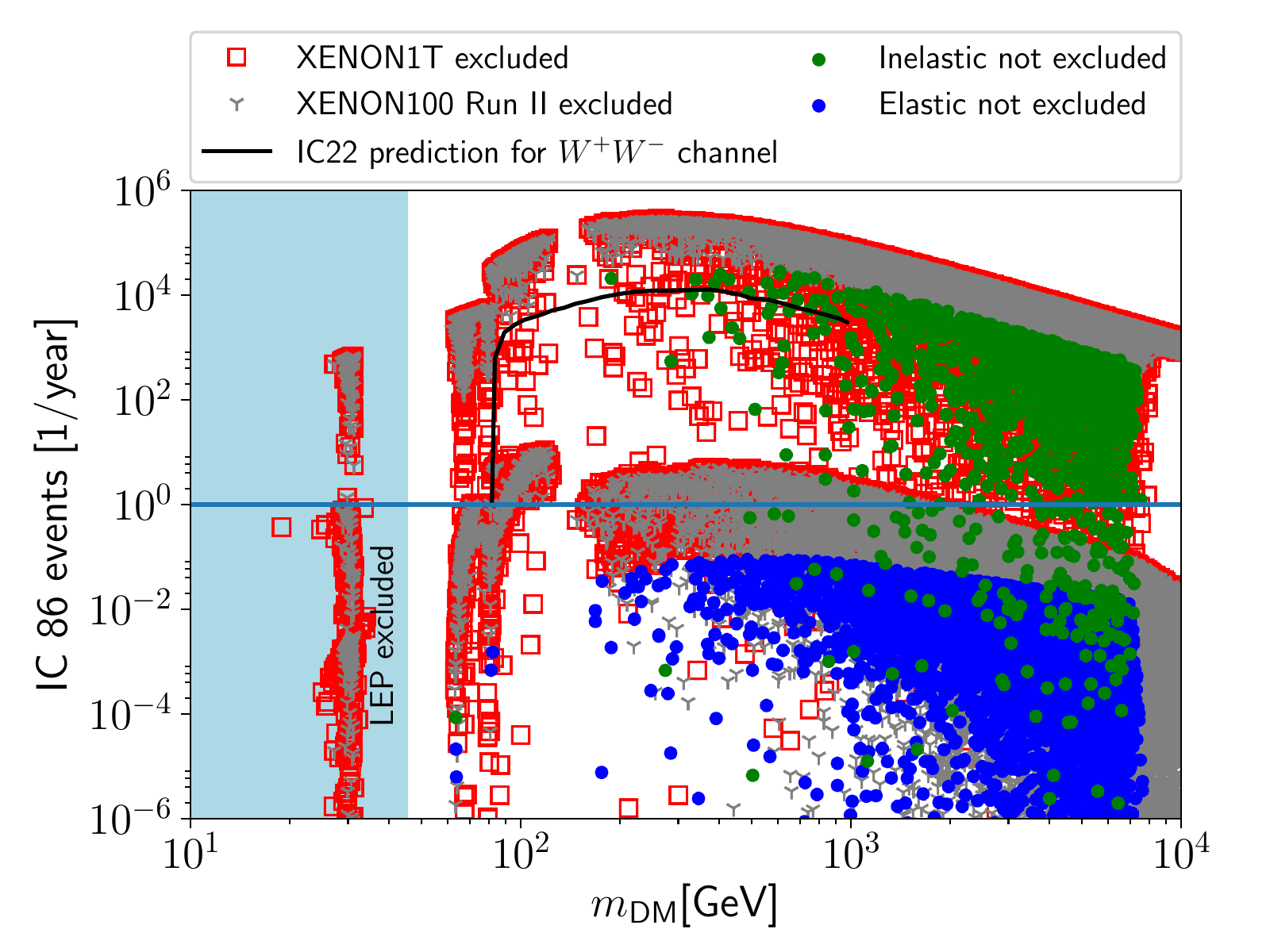}
 \caption{The expected number of events per year in the current \textsc{IceCube} configuration with 86 strings (\text{IC86}) as a function of the DM mass from inelastic and elastic DM scattering in the Sun. The black line marks the number of expected events in the $W^+W^-$ channel for an \text{IC22} study \cite{Shu:2010ta}. The blue line marks one event per year for orientation. Also shown are the points excluded by \textsc{XENON1T} \cite{Aprile:2018dbl}, \textsc{XENON100} \cite{Aprile:2011ts,Aprile:2017aas} and LEP from the invisible $Z^0$ boson width \cite{Cao:2007rm,Lundstrom:2008ai}. In the lower plot, coannihilation processes are enhanced by the small scalar-fermion mass difference.}
 \label{fig: DM mass vs SI,Events}
\end{figure}

In Fig.\ \ref{fig:6}, we plot all scotogenic model points that survive Higgs mass, DM relic density, neutrino mass/mixing and LFV violation constraints in the plane DM mass and neutral scalar mass splitting. The points are color coded for expected IC86 events (temperature scale) and exclusion of elastic scattering by \textsc{XENON1T} (red boxes) \cite{Aprile:2018dbl}. Following the observations by \textsc{DAMA/LIBRA} \cite{Bernabei:2018jrt,Bernabei:2019ajy}, several other experiments have specifically searched for inelastic DM. In a first analysis of 100.9 live days of data with a fiducial volume containing 48 kg of liquid xenon, \textsc{XENON100} (grey shaded areas) excluded the high mass (iodine) preferred region with mass splittings up to 140 keV \cite{Aprile:2011ts}. The Run II data with 224.6 live days of data with a fiducial volume containing 34 kg of liquid xenon was later reanalyzed in the context of effective field theory (EFT) using the correspondence
\begin{equation}
 \sigma_N^0=(C_1^N)^2{\mu_N^2\over\pi},
\end{equation}
where $\mu_N$ is the DM-nucleon reduced mass and $C_1^N$ is the coefficient of the spin-independent operator \cite{Aprile:2017aas}. Since in our case the interaction is not isospin-conserving, we translate the limits on $C_1^N$ from Ref.\ \cite{Aprile:2017aas} to the proton cross section as \cite{Yaguna:2016bga}
\begin{equation}
 \sigma_p^0=\sigma_N^0 \left[{Z\over A}+\left(1-{Z\over A}\right){b_n\over b_p}\right]^{-2},
\end{equation}
assuming for simplicity $A=132$ for the xenon isotope with the largest abundance. This excludes the mass region above 300 GeV with mass splittings up to 250 keV (full black lines). \textsc{PandaX-II}, who analyzed 79.6 live days of data with a fiducial volume containing 329 kg of liquid xenon, presented their results only for fixed DM masses of 1 and 10 TeV \cite{Chen:2017cqc}. When interpolated and translated for isospin violation, they give a similar, but slightly weaker exclusion curve (dashed black lines) as/than \textsc{Xenon100} due to the smaller recoil energy window and larger background. The exclusion of larger mass splittings with direct detection experiments is limited by the maximum recoil energy and would require much larger cross sections than expected from electroweak interactions \cite{Bramante:2016rdh}. The low-mass (sodium) point (black stars), to which a good fit is still possible, albeit with a large cross section of $10^{-2}$ pb and different isospin violation $b_n/b_p\simeq-0.7$ \cite{Baum:2018ekm,Kang:2018qvz}, as well as the (already excluded) high-mass (iodine) point to the \textsc{DAMA/LIBRA} signal \cite{Kang:2019uuj} are under intense scrutiny by the \textsc{DM-Ice17} \cite{deSouza:2016fxg}, \textsc{COSINE-100} \cite{Adhikari:2019off,Kang:2019fvz}, \textsc{SABRE} \cite{Antonello:2020xhj} and \textsc{ANAIS-112} \cite{Amare:2021yyu} experiments, which are expected to provide a 3$\sigma$ C.L.\ test of this signal by autumn 2022.

\begin{figure}
 \centering
 \includegraphics[width=0.6\textwidth]{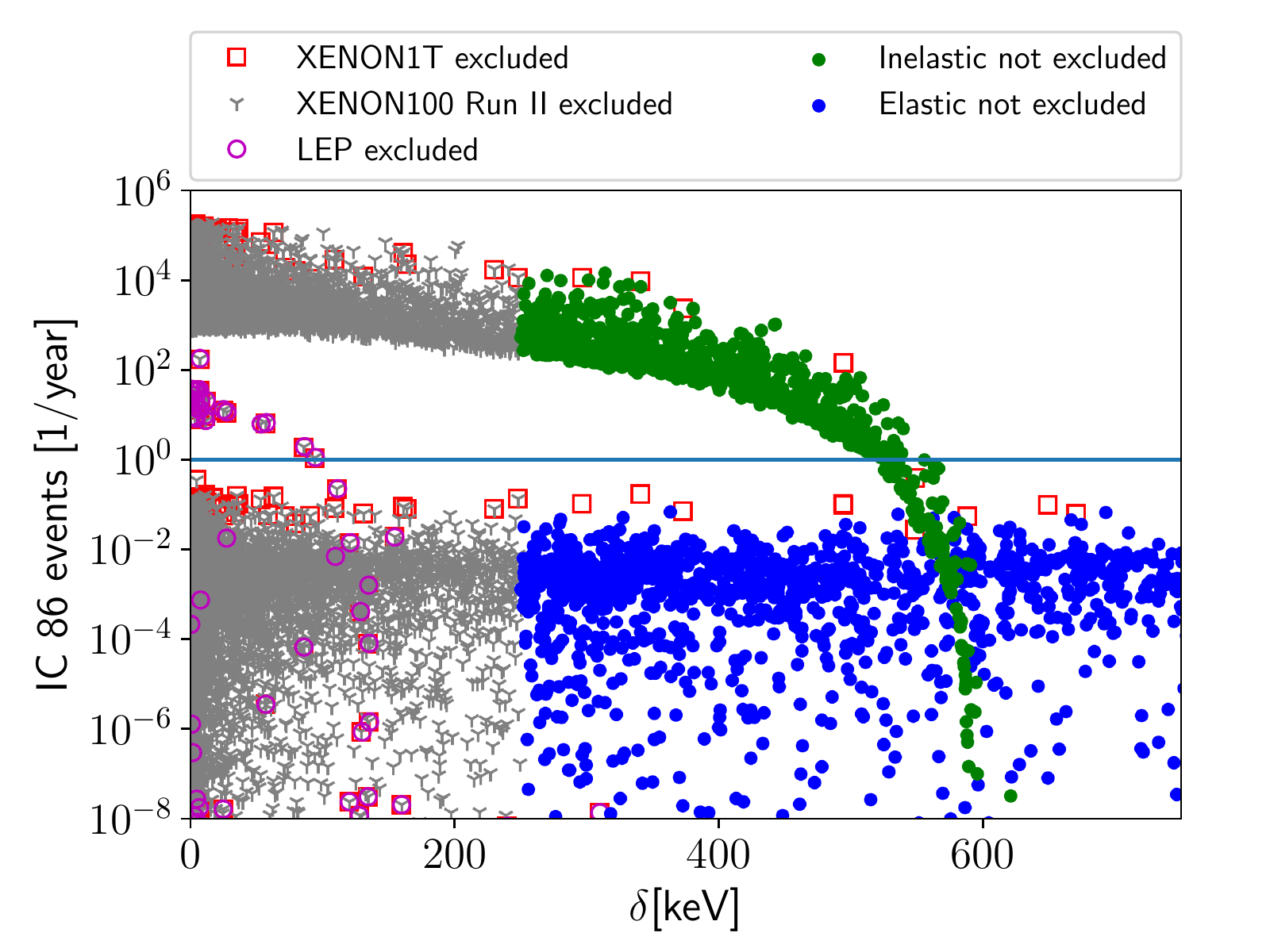}
 \includegraphics[width=0.6\textwidth]{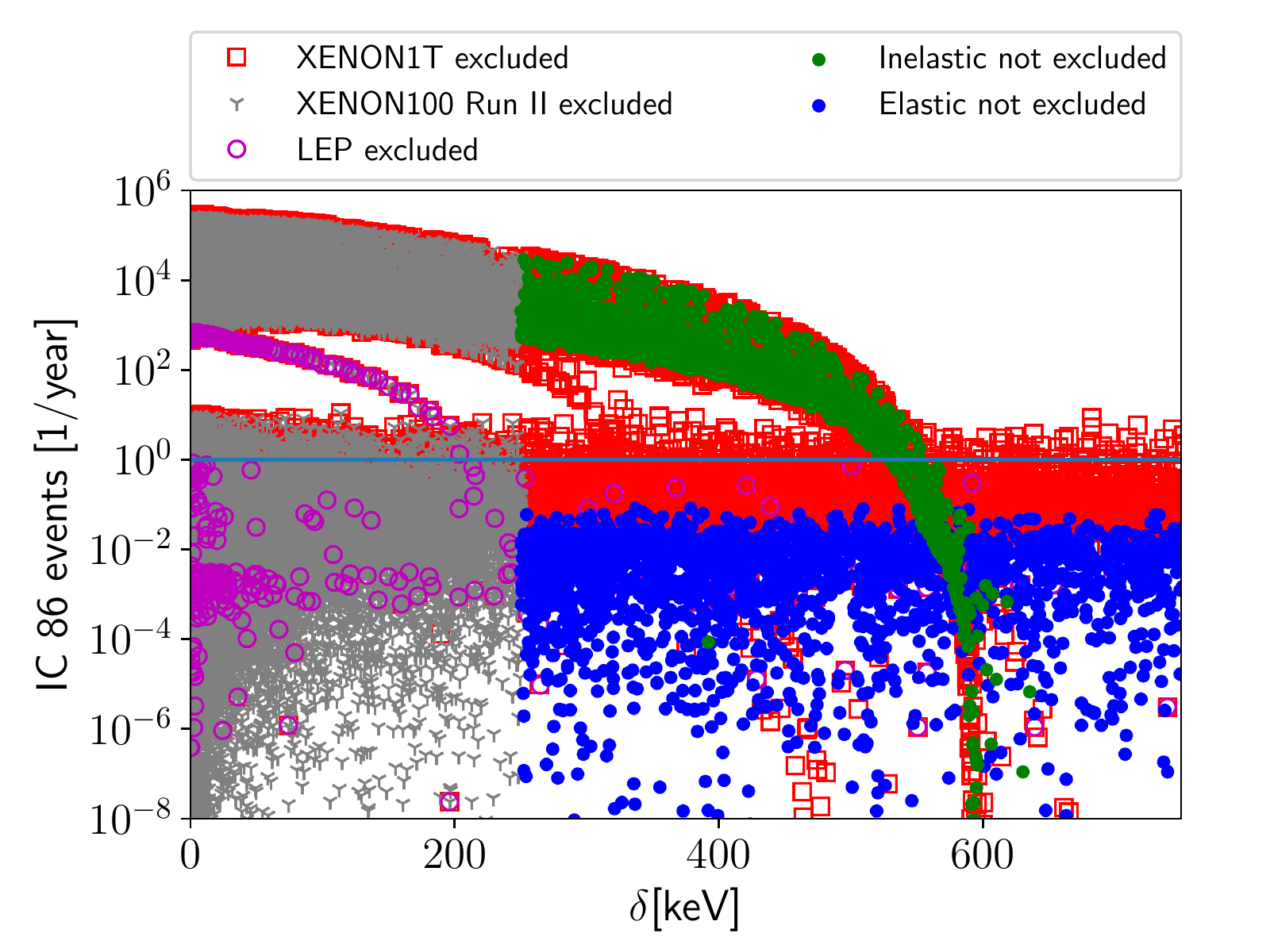}
 \caption{Same as Fig.\ \ref{fig: DM mass vs SI,Events} as a function of the neutral scalar coupling splitting $\delta$.}
    \label{fig: lam5 delta vs Events}
\end{figure}
\begin{figure}
 \centering
 \includegraphics[width=0.6\textwidth]{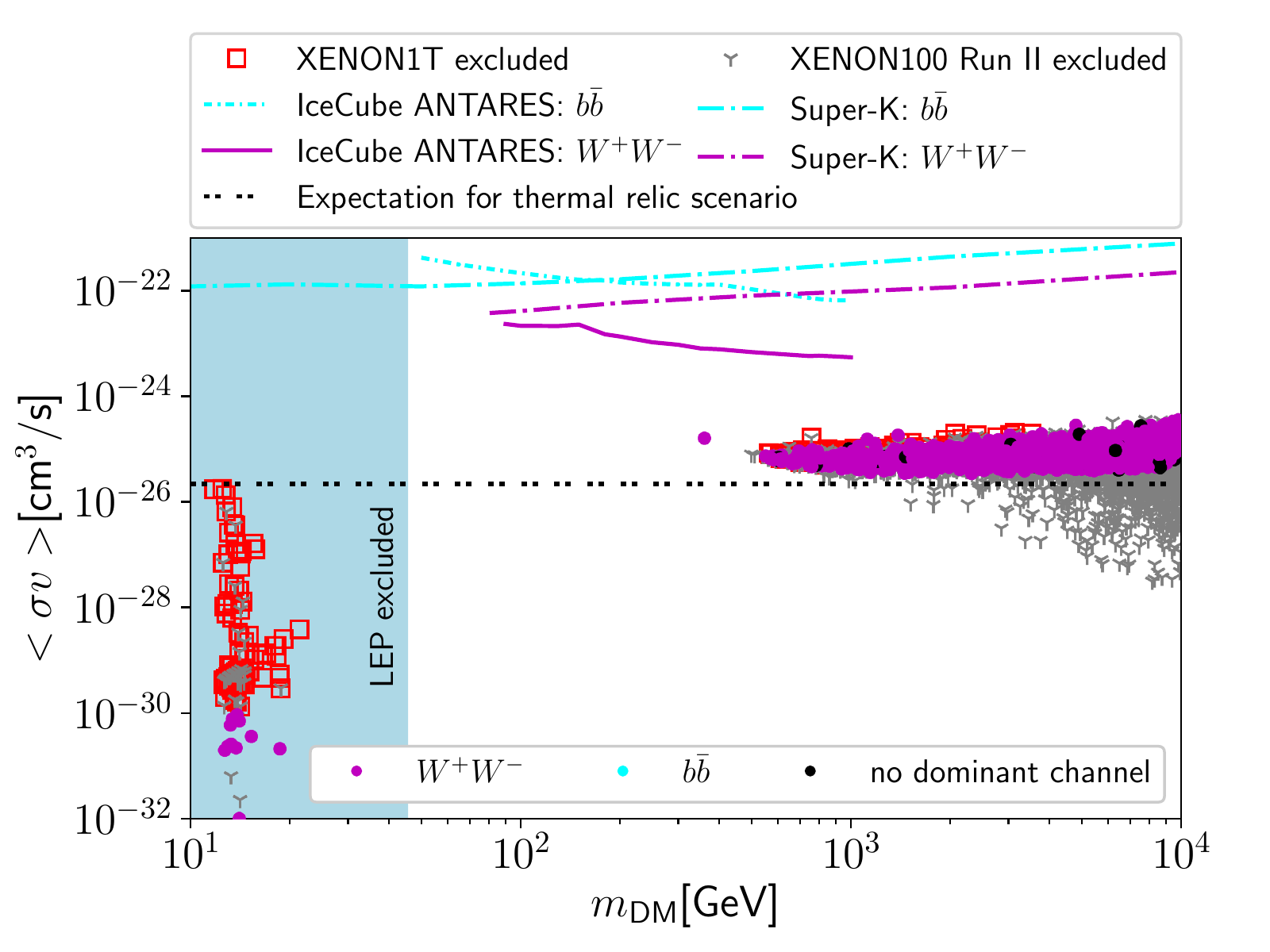}
 \includegraphics[width=0.6\textwidth]{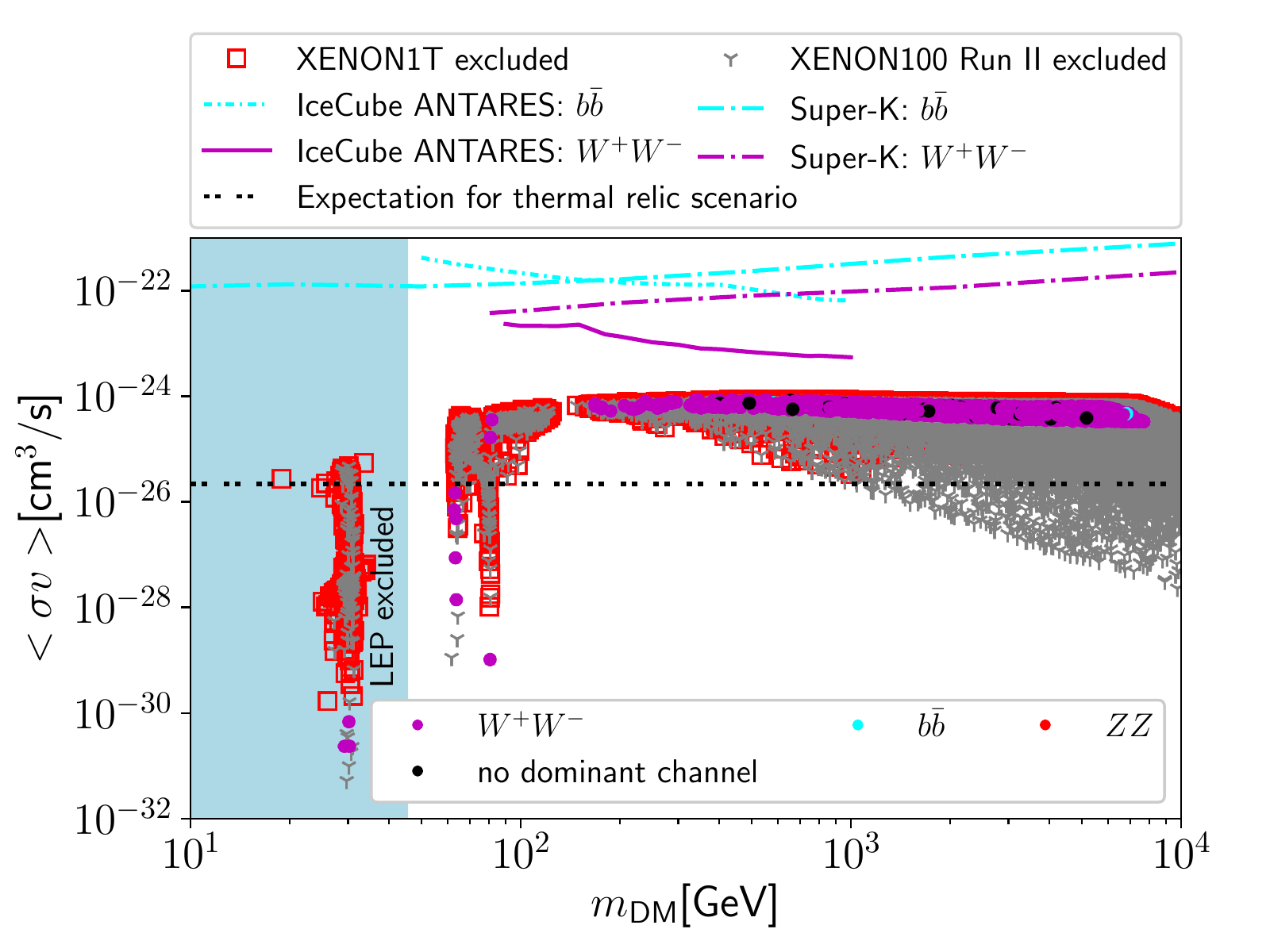}
 \caption{Thermally averaged cross section $\langle \sigma v \rangle$ with combined \textsc{IceCube} \textsc{ANTARES} \cite{Aartsen:2020tdl} and \textsc{Super-Kamiokande} \cite{Abe:2020sbr} exclusion limits as a function of the DM mass. All points are colored according to the main annihilation channel, provided there is one with a branching ratio of over 50\%. Also shown are the points excluded by \textsc{XENON1T} \cite{Aprile:2018dbl}, \textsc{XENON100} \cite{Aprile:2011ts,Aprile:2017aas} and LEP from the invisible $Z^0$ boson width \cite{Cao:2007rm,Lundstrom:2008ai} as well as the expected cross section for a thermal relic \cite{Steigman:2012nb}. In the lower plot, coannihilation processes are enhanced by the small scalar-fermion mass difference.}
    \label{fig:sigma v}
\end{figure}

\subsection{Expected \textsc{IC86} event rates from (in)elastic DM scattering in the Sun}

The expected number of events in the current \textsc{IceCube} configuration with 86 strings (\text{IC86}) was already shown by color coding in Fig.\ \ref{fig:6}. It ranges from less than $10^{-6}$ to more than $10^5$ per year. For both the random and the coannihilation scan, at least ten events are expected for neutral scalar mass splittings $\delta\leq(500\pm20)$ keV. Points below 250 keV were already excluded by \textsc{XENON100} \cite{Aprile:2017aas}, but half of the parameter space could be tested with \textsc{IC86} for the first time. From Eq.\ (\ref{eq: del lam5}), the non-observation of the predicted events would set a lower bound on
\begin{equation}
  \lambda_5 \gtrsim 1.6 \cdot 10^{-5} \cdot m_{\rm DM}/{\rm TeV}.
\end{equation}
In the following, we analyze the expected event rates in more detail. Fig.\ \ref{fig: DM mass vs SI,Events} shows the expected number of events per year as a function of the DM mass for both inelastic and elastic DM scattering in the Sun, i.e.\ when both occurs two points are shown. Points excluded by \textsc{XENON1T} \cite{Aprile:2018dbl}, \textsc{XENON100} \cite{Aprile:2011ts,Aprile:2017aas} and LEP from the invisible $Z^0$ boson width \cite{Cao:2007rm,Lundstrom:2008ai} are also shown. The black line shows the estimate of the number of expected inelastic events in the $W^+W^-$ channel from a previous study that had adopted the \text{IC22} configuration \cite{Shu:2010ta}, which scales roughly as expected with our \text{IC86} predictions. Note that this and other previous analyses of indirect detection of inelastic DM were motivated by the high-mass (iodine) \textsc{DAMA/LIBRA} best fit point with $\delta=(125\pm25)$ keV \cite{Nussinov:2009ft,Menon:2009qj,Shu:2010ta}. In our Fig.\ \ref{fig: DM mass vs SI,Events}, the blue line marks one event per year for orientation. As one can see, the expected event rates for models with only elastic scattering stay below this line. In the lower plot, where coannihilation processes are enhanced by the small scalar-fermion mass difference, models with larger elastic scattering rates are already excluded by direct detection experiments. Nevertheless, inelastic scattering allows for a large region of the scalar DM parameter space with and without coannihilation to be tested by \text{IC86}.

Fig.\ \ref{fig: lam5 delta vs Events} shows the expected number of events per year as a function of the neutral scalar mass splitting $\delta$. Although we scan values above $\lambda_5\geq10^{-10}$, the neutrino masses constrain $\lambda_5$ to be larger than $10^{-9}$. In addition, in the coannihilation scenario the relic density mostly requires values above $10^{-8}$. From Fig.\ \ref{fig:6} we know that direct detection experiments exclude neutral scalar mass splittings below 250 keV, which for $m_{\rm DM} \geq500$ (200) GeV in the normal (coannihilation) scan translates through Eq.\ (\ref{eq: del lam5}) to a limit of $\lambda_5\gtrsim4.1\ (1.7)\ \cdot10^{-6}$. With \textsc{IC86}, mass splittings up to 500 keV and at least two more orders of magnitude in $\lambda_5$ could be tested up to $1.6\cdot10^{-4}$ (or beyond) for $m_{\rm DM}=10$ TeV (or larger). While the event rate falls quickly for inelastic scattering towards the kinematic edge, it is of course independent of both $\delta$ and $\lambda_5$ in the elastic case.

\subsection{Limits from DM annihilations in the Galactic Center}

Neutrino telescopes are also used to set bounds on the self annihilation of DM in the Galactic Center. We test the scotogenic model with the limits on the thermally averaged self annihilation cross section $\langle\sigma v\rangle$ set by a joint analysis of \textsc{ANTARES} and \textsc{IceCube} \cite{Aartsen:2020tdl}, assuming a NFW halo profile \cite{Navarro:1995iw}. The results are shown in Fig.\ \ref{fig:sigma v}. All points are color coded for their dominant annihilation channel (i.e.\ $b\bar{b}$ quark, $W^+W^-$ and $ZZ$ boson pairs). Points with no branching ratio larger than $50\%$ are marked as ``no dominant channel''. The predictions agree roughly with the naive expectation for a thermal relic \cite{Steigman:2012nb}, but can be considerably larger in the coannihilation scenario, where the coannhiliation processes increase rather than decrease the predicted DM relic density \cite{KRRYZ13}. Contrary to DM annihilation in the Sun, the direct detection experiments \textsc{XENON1T} \cite{Aprile:2018dbl} and \textsc{XENON100} \cite{Aprile:2011ts,Aprile:2017aas} now exclude points with {\em lower} expected rates, while {\em larger} rates remain viable. This can be attributed to the fact that the DM density in the Galactic Center is now fixed by the NFW profile and not determined by (in)elastic scattering and the capture rate. Low-mass points annihilating mostly into $b\bar{b}$ quarks are already excluded by the direct detection experiments and LEP from the invisible $Z^0$ boson width \cite{Cao:2007rm,Lundstrom:2008ai}, whereas the limits from \textsc{Super-Kamiokande} \cite{Choi:2015ara} are much weaker. For high-mass points, DM annihilation into $W^+W^-$ bosons is dominant. While the points from our random scan lie still considerably below the limits set by \textsc{ANTARES} and \textsc{IceCube} \cite{Aartsen:2020tdl}, those from our coannihilation scan are less than an order of magnitude smaller and should be within reach of future \textsc{IC86} analyses. It would be particularly interesting to extend the mass range of this analysis from 1 to 10 TeV and beyond.

\section{Summary and outlook\label{sec: Summary}}

To summarize, we have investigated in this paper the indirect detection prospects of scalar DM in the scotogenic model. We have focused on DM annihilation into neutrinos in the Sun, but also in the Galactic Center. After a brief review of the particle content, interactions and neutrino mass generation in the scotogenic model, we described in detail the elastic and inelastic DM scattering processes induced by Higgs and $Z$ boson exchanges in the Sun, which determine the WIMP capture rate and thus also the annihilation rate into neutrinos and particles that decay into them. We then implemented the capture rate from inelastic scattering in \textsc{DarkSUSY 6.2.3} and interfaced it with \textsc{micrOMEGAs 5.0.8} and our updated routine for the expected neutrino fluxes on Earth and the effective area of the \textsc{IceCube} detector in its current 86-string configuration.

We then performed two large numerical scans spanning the theoretically allowed parameter space, i.e.\ a random scan and one with enhanced scalar-fermion coannihilation from a small DM-sterile neutrino mass splitting, which is known to increase the relic density and extend the viable scalar DM mass region from above 500 GeV to above 200 GeV. Experimental constraints were imposed from the known neutrino mass differences and mixing angles, LFV, the searches for new neutral and charged scalars at LEP and LEP2, the LHC measurements of the Higgs boson mass and invisible width, and from direct and previous indirect DM searches. 

First, we found that direct, but not indirect detection experiments constrain the spin-independent elastic scattering cross section, in particular in the coannihilation scenario. We also found that a considerable fraction of the models lie below the neutrino floor, which may render direct detection difficult. We then showed that direct detection experiments cover only half of the parameter space for inelastic scattering, i.e.\ inelasticities up to 250 keV. The higher kinetic energy of DM in the Sun therefore leaves ample room for a dedicated analysis with \textsc{IC86}, that would cover inelasticities up to at least 500 keV. The expected rates extend well beyond 10$^3$ per year. The non-observation of the predicted neutrino events would translate into a lower limit on the scalar coupling $\lambda_5\gtrsim1.6\cdot10^{-5}\cdot m_{\rm DM}/$TeV. For larger couplings, only elastic scattering has to be considered. In this case, the expected event rates for models that are not yet excluded by the direct detection experiments do not exceed 0.1 per year. We reminded the reader that the coupling $\lambda_5$ has to be naturally small, since if it was exactly zero, the neutrinos would be massless and lepton number would be conserved, leading to a larger symmetry of the Lagrangian.

Models with elastic and inelastic scattering can also be tested with DM annihilation in the Galactic Center, assuming e.g.\ a NFW DM profile. Here, we found that direct detection experiments exclude mostly models with lower thermally averaged cross sections. This could be attributed to the fact that the DM density in the Galactic Center was fixed by the NFW profile and not determined by (in)elastic scattering and the capture rate. Low-mass points annihilating mostly into $b\bar{b}$ quarks were already excluded by direct detection experiments and LEP, whereas the limits from \textsc{Super-Kamiokande} were much weaker. For high-mass points, DM annihilation into $W^+W^-$ bosons was dominant. There, a previous combined analysis by \textsc{ANTARES} and \textsc{IceCube} led to limits that were two orders of magnitude above our predictions. For the coannihilation scenario, our predictions are, however, less than an order of magnitude smaller and thus within reach of future \textsc{IC86} analyses, in particular for TeV-scale DM.

Our results generalize to models with several scalar multiplets where the mass splitting between the neutral components is small. A particularly interesting case for future study would be the AMEND model with small singlet-triplet scalar mass splitting, which could again be small due to an otherwise larger symmetry of the Lagrangian \cite{Farzan:2010mr}.

\section*{Acknowledgments}

We thank 
Carsten Rott for useful comments on the manuscript.
This work has been supported by the DFG through the Research Training Network 2149 ``Strong and weak interactions - from hadrons to dark matter'' and by BMBF through Verbundforschung grant 05A20PM2.


\providecommand{\href}[2]{#2}\begingroup\raggedright\endgroup

\end{document}